\RequirePackage{fix-cm}

\documentclass[smallcondensed]{svjour3}     % onecolumn (ditto)

\smartqed  
\usepackage{multirow}
\usepackage{graphicx}

\usepackage{mathptmx}      % use Times fonts if available on your TeX system
\usepackage[round]{natbib}
\usepackage{comment}
\usepackage{latexsym}

\usepackage{amsmath,amssymb}
\usepackage{commath}

\usepackage[caption=false]{subfig}

\usepackage{nameref,hyperref}

\usepackage{xspace}
\usepackage{hyperref}
\usepackage{multicol}
\usepackage{balance}

\AtBeginDocument{%
  \providecommand\BibTeX{{%
    \normalfont B\kern-0.5em{\scshape i\kern-0.25em b}\kern-0.8em\TeX}}}

\usepackage{url}
\usepackage{hyperref}
\usepackage{graphicx}
\usepackage{dcolumn}
\usepackage{xspace}
\usepackage{balance}
\usepackage{bm}
\usepackage{subfig}

\usepackage{array}
\usepackage{tabularx}

\newcolumntype{P}[1]{>{\raggedright\arraybackslash}p{#1}}
\newcolumntype{Q}[1]{>{\raggedleft\arraybackslash}p{#1}}

\newcommand{\etal}{{\emph{et al.}}\xspace}

\newcommand{\taseg}{{\sc BlockFix}\xspace}
\newcommand{\taparse}{{\sc FragFix}\xspace}
\newcommand{\tatype}{{\sc TypeFix}\xspace}

\usepackage{color, colortbl}
\definecolor{Gray}{gray}{0.9}

\usepackage{ifthen}
\usepackage{amssymb}
\newboolean{showcomments}
\setboolean{showcomments}{true} % toggle to show or hide comments
\ifthenelse{\boolean{showcomments}}
  {\newcommand{\nb}[2]{
    \fcolorbox{gray}{yellow}{\bfseries\sffamily\scriptsize#1}
    {\sf\small$\blacktriangleright$\textit{#2}$\blacktriangleleft$}
   }
   
  }
  {\newcommand{\nb}[2]{}
   
  }

\usepackage{siunitx}
\usepackage{soul}

\newcounter{RQCounter}
\newcounter{HCounter}
\newcounter{RSCounter}
\usepackage{framed}

\usepackage{quoting}
\quotingsetup{leftmargin=10pt,rightmargin=10pt,vskip=3pt}

\usepackage{enumitem} 
\setlist[itemize]{noitemsep}
\setlist[itemize]{nolistsep}

\usepackage{listings} %code highlighter
\usepackage{color} %use color
\definecolor{mygreen}{rgb}{0,0.6,0}
\definecolor{mygray}{rgb}{0.5,0.5,0.5}
\definecolor{mymauve}{rgb}{0.58,0,0.82}
 
\definecolor{darkgray}{rgb}{.4,.4,.4}
\definecolor{purple}{rgb}{0.65, 0.12, 0.82}
 
\lstdefinelanguage{JavaScript}{
keywords={typeof, new, true, false, catch, function, return, null, catch, switch, var, if, in, while, do, else, case, break},
keywordstyle=\color{black}\bfseries,
ndkeywords={class, export, boolean, throw, implements, import, this},
ndkeywordstyle=\color{darkgray}\bfseries,
identifierstyle=\color{black},
sensitive=false,
comment=[l]{//},
morecomment=[s]{/*}{*/},
commentstyle=\color{purple}\ttfamily,
stringstyle=\color{red}\ttfamily,
morestring=[b]',
morestring=[b]"
}
 
\usepackage[dvipsnames]{xcolor}

\usepackage{amsmath}
\usepackage[noend]{algpseudocode}
\usepackage{varwidth}

\usepackage[commentsnumbered,linesnumbered]{algorithm2e}
\makeatletter
\def\BState{\State\hskip-\ALG@thistlm}
\makeatother

  {\list{}{\leftmargin=#1\rightmargin=#1}\item[]}%
  {\endlist}

\newenvironment{smcodetabbing}
  {\relax\ttfamily\small\tabbing}
  {\endtabbing}

\newcommand{\premdone}[1]{}

\newcommand{\sofo}{{\small {\sc StackOverflow}}\xspace}

\newcolumntype{L}[1]{>{\raggedright\let\newline\\\arraybackslash\hspace{0pt}}m{#1}}
\newcolumntype{C}[1]{>{\centering\let\newline\\\arraybackslash\hspace{0pt}}m{#1}}
\newcolumntype{R}[1]{>{\raggedleft\let\newline\\\arraybackslash\hspace{0pt}}m{#1}}

\begin{document}

\title{Learning Lenient Parsing \& Typing via Indirect Supervision}

\author{Toufique Ahmed \and
         Premkumar Devanbu \and %etc.
        Vincent Hellendoorn
}

\institute{Toufique Ahmed\at
              Department of Computer Science, University of California, Davis, CA, USA \\
              \email{tfahmed@ucdavis.edu}           %  \\
           \and
            Premkumar Devanbu \at
              Department of Computer Science, University of California, Davis, CA, USA \\
              \email{ptdevanbu@ucdavis.edu} 
           \and
           Vincent Hellendoorn    \at
           	Department of Computer Science, University of California, Davis, CA, USA \\
              \email{vhellendoorn@ucdavis.edu } 
}

\date{Received: date / Accepted: date}

\maketitle

\begin{abstract}

Both professional coders and teachers frequently
deal with imperfect (fragmentary, incomplete, ill-formed) code. Such fragments are
common in \sofo; students also frequently produce ill-formed code, for which instructors,
TAs (or students themselves) must find repairs.
In either case, the developer experience could be greatly improved if such code
could somehow be parsed \& typed; this makes such code more amenable to use within IDEs and allows early detection and repair of potential errors.
We introduce a \emph{lenient} parser, which can parse \& type fragments, even 
ones with simple errors. 
% Although we humans are usually only taught
%the syntactic processing \& typing of \emph{perfect}
%code, we intuitively and gradually learn to deal with \emph{imperfect}
%code. 
Training a machine learner to leniently
parse and type imperfect code requires a large
training set including many pairs of imperfect code and its repair (and/or type information); such training sets are limited by human effort and curation.
In this paper, we present a novel, indirectly supervised, approach to 
train a lenient parser, 
%learn how to \emph{leniently} parse \& type imperfect code, 
without access to such human-curated training data. 
We leverage the huge corpus of \emph{mostly correct} code
available on Github, and the massive, efficient learning capacity of Transformer-based NN architectures. 
Using GitHub data, we first create a large dataset of 
fragments of code and corresponding tree fragments and type annotations; we then randomly corrupt
the input fragments (while requiring correct output) by seeding errors
that mimic corruptions found in \sofo and student data.
Using this data, we train high-capacity transformer models to overcome both fragmentation and corruption. With this novel approach, we can achieve reasonable performance on parsing \& typing \sofo fragments; we also demonstrate that our approach performs well on shorter student error program and achieves best-in-class performance on longer programs that have more than 400 tokens. We also show that by blending Deepfix and our tool, we could achieve 77\% accuracy, which outperforms all previously reported student error correction tools.%a large dataset of student errors.


\end{abstract}

%!TEX root=main.tex

\section{Introduction}
Most of the development tools require syntactically correct, well-typed code; 
the rest will usually be rejected in some fashion
by static or dynamic checks within the tool. 
However, developers often have to confront and work with
fragmentary, malformed code. Two prominent settings of concern are
a) partial, or flawed, code fragments from \sofo, and b) malformed code in 
student assignments. \sofo fragments are often useful, but may 
not be syntactically complete and correct. Likewise, learners struggle with syntax~\citep{mccracken2001multi},
and frequently make mistakes; the diagnosis and repair of syntax errors can be quite
a challenge, especially for beginners. TAs and professors then have to expend valuable contact hours
helping students repair such mistakes. 
Yet such fragmentary code is often already ``mostly correct'', requiring at most a few corrections; hence, it is not unrealistic to consider automating this process \citep{WangRepair,gupta2017deepfix}.

Given the demonstrated success of machine learning at similar tasks in other domains
(e.g., fixing errors in writing) there is good prior motivation to attempt several relevant tasks
here: a) \emph{Leniently parse} \sofo fragments, so that properly-constructed abstract syntax
tree (AST) fragment can be created, even from malformed/partial fragments, and made
available for  use by an IDE, b) \emph{Leniently parse} malformed student code, 
while locating and fixing errors therein.  c) \emph{Leniently type-annotate} such problematic code fragments, 
providing further information to IDEs to add necessary glue code (declarations, imports, \emph{etc}).
To our knowledge this final lenient typing task above
has not been previously attempted, for malformed code fragments. 
%We provide a unified, indirect-supervision approach
%that trains a  learner that achieves state-of-the art performance for all 3 tasks. 

%In this paper, we propose an integrated approach to all 3 of the tasks above, the
%does not rely on any human-created data. This approach was motivated the observation
%that as humans learning to  coder, we are primarily presented
%with complete, properly constructed code examples, and then also taught
%the details correct syntactic constructions, proper type annotations and sound type-inference
%regimes. Armed with this knowledge,  we learn on-the-fly to repair
%incomplete and malformed programs; this skill helps us manually overcome erroneous
%and incomplete constructions that frequently occur in \sofo, and also help those less
%experienced than us with programming find and fix their syntax errors. 

%So the question naturally arises; can we develop a new training regime that ameliorates
%the need for human effort to create (or curate) training data? 
We develop a novel approach to parsing and typing that relies on indirect-supervision training,
using only (mostly) correct code taken from Github. This code (mostly)
compiles and is thus usually syntactically correct, and is well-typed. This code
is easily processed by (\emph{e.g.}) Eclipse JDT  to yield massive volumes of matched
tuples of source, ASTs, and type annotations. We take this matched data,
and abuse the input source code in various ways to create challenging training data,
while leaving the (desired) ASTs and types alone.  First,
we chop it up randomly to create fragments (with matching types and ASTs) that mimic
the kinds of fragments found in \sofo. Second, we
randomly corrupt it (while retaining correct AST and types on the desired output) to reflect
the repair of typical errors found in \sofo fragments and in student code. We use this challenging data
to train a high-capacity
neural network to leniently parse and type imperfect, fragmentary code, by forcing it to minimize
its loss against the desired, correct output. 
To summarize: 
\premdone{Say something about indirect supervision}
\begin{enumerate}
\item 
We use an indirect-supervision
approach, which leverages GitHub code repos to create
massive amounts of ``incorrect-fixed" training pairs, 
without relying on human annotation. 
We use this data to train high-capacity, efficient  neural Transformer architectures, 
to leniently fix, parse and type fragmentary and incorrect
code. 
\item
We use a 2-stage approach, with two different neural networks, one of which learns to model (and fix)
block nesting structure, and the other which learns to model (and fix) fragments of code. This
combination allows us to deal with very long-distance
syntactic dependencies within a sequence-based neural network, and thus improve
performance on our parsing tasks. 
%\item 
%Our training approach enables our model to learn to parse, fix, and type
%fragmentary, incorrect code within the same, unified scheme. 
\item
Compared to earlier algorithmic work on robust parsers, 
our approach is fairly \emph{language-agnostic}: we make minimal assumptions about
the language, except for the existence of a parser and static typer to create training data. To port
to another language also requires identification of block delimiters, expression delimiters, and
statement delimiters. (respectively, {\tt '\{\}', '()', ';'}), as will be clear below. 

\item
We have  evaluated our approach using a combination of automated
and manual protocols, and demonstrate that we achieve good  performance on the novel
typing task, and improve upon a prior baseline for repairing student code, for longer fragments. 
We explicitly compare our tool with DeepFix~\citep{gupta2017deepfix} and the tool proposed by Santos \etal~\citep{santos2018syntax}. 
\item
We have released our data, to the extent permissible (for student data)\footnote{It's possible for others to license the data, however, as did we.}, and made
our implementation available. 
\end {enumerate}

We also point out that our approach could be used as a pre-training adjoint to existing translation
based approaches, which rely on human-created datasets; thus in addition to improving on 
prior performance, our indirect supervision approach could be supplemented with direct supervision to
yield further improvements. 

The remainder of the paper is organized as follows. 
Section \ref{motive} presents the motivation and background of our research.
Section \ref{ta} discusses the technical approach of the paper. 
Section \ref{eval} presents the results of this research. 
Section \ref{relatedwork} describes some prior research relevant to our work. 
Section \ref{sec:discussion} discusses the implications of the work and provides some future direction.  
Finally, Section \ref{sec:conclusion} concludes the paper.

%!TEX root=main.tex

\section{Motivation \& Background}
\label{motive}
\sofo  is now the preferred source of coding examples for developers. Given any coding quandary 
about core language features, or specific APIs, one can find answers, with illustrative
code examples, on \sofo. However, the code examples are often
fragmentary: just a few stand-alone lines of code, which 
are not complete, parseable units of Java code. 
If these fragments could  be parsed into an AST\footnote{AST = Abstract Syntax Tree} form, and also typed, then it would be much
easier to paste them into an IDE: the IDE could assist by adding {\small\tt import} statements
to import packages relevant to the types used in the fragments, adding declarations for needed variables, suggest renaming
of variables occurring in the fragment to relevant variables of corresponding types
currently in scope, and so on.

But how can ASTs and types be obtained for partial fragments of code?
Typing fragments is rarely possible, as they usually don't provide the necessary import
statements and declarations to allow the types of variables in code fragments to be inferred. 
Parsing fragments to derive ASTs is non-trivial as well. Consider an otherwise correct fragment (from the \href{https://stackoverflow.com/questions/6200533/set-textview-style-bold-or-italic}{Android section} of \sofo):

\begin{smcodetabbing}
textView.setTypeface(textView.getTypeface(), Typeface.BOLD);
\end{smcodetabbing}

\noindent For such a well-constructed 
fragment, one can simply wrap the fragment in a dummy method, and invoke a parser, which would
provide an AST for the entire dummy method, from which one can easily extract the parse for just the
fragment. Although this \emph{``wrap-and-parse''} trick is simple and appealing, we estimate based on a manual examination of 200 randomly sampled fragments, that 28\%
of \sofo fragments\footnote{28\% is a point estimate. The 95\% Wald confidence interval, on a binomial
estimator with a sample size of 200, is 22-35\%.} are not parseable due to various kinds of coding errors. Such fragments are quite common on \sofo, often missing delimiters, including  ellipses,  and
missing declarations;  the following example from Stackoverflow is fairly typical \footnote{https://stackoverflow.com/a/54596387}, and cannot be parsed by Eclipse JDT.

\begin{smcodetabbing}
{\tt Optional<String> getIfExists() \{ }\\
{\tt   		\qquad $\ldots$}\\
{\tt  	    \qquad return Optional.empty();}\\
{\tt\}}\\
\end{smcodetabbing}

%\begin{smcodetabbing}
%{\tt byte[] bytes = \{ $\ldots$ \} }\\
%{\tt String str = new String(bytes, "UTF-8");}\\
%\end{smcodetabbing}

\noindent These fragments resist processing via the simple \emph{``wrap-and-parse''} trick, and require more
intelligent handling. Table~\ref{parse} presents a few more \sofo fragments (along with this example) and the parse trees generated by our approach.

A more intelligent, lenient parsing approach could have benefits beyond \sofo. Student code, for example 
is rife with similar, syntactic errors; automatically fixing these could be very helpful  for teachers and learners. 
In our experiment, we use the Blackbox dataset~\citep{brown2014blackbox}
which contains millions of examples of student submissions from around a million users. 
Fig~\ref{jsorig} is an example from 
this dataset. Both Eclipse and IntelliJ IDEs fail to repair this example because none of the IDEs have any hand-crafted rule to solve this problem.

 %~\cite{brown2014blackbox} which we describe below in Sec.~\ref{sec:studentcode}. 
%\prem{An example from student code, with misbalanced curlies and other nonsense. }
%\toufique{An example added. I did not add any explanation. Could not highlight the error in line 6 and 9. I took it from blackbox. It already had error at line 6 (missing close paren) and I added one extra brace at 9. Tested this one. Our model can repair it}
\vspace{-0.1in}
\begin{figure}[htb]
\captionsetup{aboveskip=-24pt,belowskip=-9pt}
\centering
\begin{lstlisting}[language=Java,basicstyle=\scriptsize\tt]
public class Modulus
{
	public static void main(String[] args){
    
    	for(int i = 0; i < 12; i = i + 2){
        	System.out.println("i = " + i);
        	System.out.println("3 - i (" + (i) + ") = " (3 - i));
    	}
   	 
	}
}

\end{lstlisting}
\caption{Incorrect (verbatim) student code sample}
\label{jsorig}
\end{figure}
\noindent Note the missing ``+'' on line 7 before ``$(3-i)$''. 
Simple syntax errors 
challenge and frustrate beginners~\citep{mccracken2001multi}.  
A lenient parser pipeline could deal with these: it can fix the error, and in the case of the student program, provide a full parse tree that indicates the context where the missing ``+'' is needed. 
%\noindent Traditional parsing methods, and even ``hacks" like wrapping within a dummy function body, fail miserably
%in such cases; typing, is of course, quite impossible.
Our overarching research goal is a kind \emph{lenient program analysis} which exploits 
powerful deep neural network models of code 
to enable IDEs to work more intelligently with malformed fragments by guessing their intended structure; 
in this paper, we focus on \emph{lenient parsing and typing} specifically aimed at managing
the vagaries of  \sofo fragments and  student code. \sofo fragments, thus rendering them a greater proportion of them more usable within an IDE. 

%\prem{is this island Parsing? no, it isn't, because the islands themselves are malformed, it wouldn't work}

%!TEX root=main.tex

\section{Technical Approach}
\label{ta}
For the lenient parser,  we use a pipeline with two learned deep-neural network (DNN) stages. 
The first DNN stage learns to repair errors in nested block structure (``\taseg''), and the 
second stage learns repair and parse syntactically incorrect fragments (``\taparse''). This two-stage approach is needed to handle long-range
dependencies, as we discuss below. The lenient typer (\tatype) is a single-stage learned model. 
All of the 3 learned models are built using Transformer-based architectures, which are explained below. 

\subsection{Overall Architecture}
We begin with the intuition that Natural Language (NL)  parsing is a helpful platform to build learned
models to process malformed code. 
NL is complex, ambiguous, and challenging to parse. 
Syntactic (``constituency'') parsing is a core NLP problem, that has been refined over decades. 
Given that code corpora have been shown to be ``natural", NLP parsing technology holds promise for lenient parsing of code. 

Traditionally, however, effective NL  parsers were tricky beasts that required a lot of algorithm engineering.
This approach changed substantially when a completely data-driven DNN
architecture~\citep{vinyals2015grammar} was shown to be remarkably effective at parsing. Rather than using a
pre-conceived formalism (\emph{e.g.,} probabilistic context-free grammars) with associated algorithms, 
they render parsing as translation. Just 
as DNNs could \emph{learn} to translate from English to German from large datasets of
aligned English-German sentence pairs, they could \emph{learn to effectively parse} from
aligned pairs of sentences and associated (serialized) parse trees.
%This remarkable approach worked very well indeed. 

To our knowledge, this learned parsing-as-translation
approach has not been used for fragmented, noisy code, but it is \emph{prima facie} well-suited.  
Unlike with NL, where parse trees (for training) must be hand-constructed by experts, large amounts
of parsed code can be freely harvested by compiling  complete projects from GitHub. 
Our \underline{\emph{core idea}} is this: 
while parsing complete files requires correct code and correct build set-up, 
the fragments of code contained therein have regularities (thanks to the well-known
naturalness~\citep{hindle2012naturalness} phenomenon) that will allow a well-trained
high-capacity DNN to learn to parse most commonly-occurring 
fragments of code,  even wrong ones, \underline{\emph{without}} the benefit of build and parsing context. 
For greater capacity, while Vinyals \etal used an older, % attention-enhanced
Sequence-to-sequence (seq2seq, with attention) recurrent-neural-network
(RNN) approach,  we use % However, we used 
a newer transformer-based model~\citep{vaswani2017attention} which is known
 to outperform older seq2seq approaches. 
Even so, there are several novel issues that arise when trying to use NL parsers for the task
of parsing fragmented and noisy code.

We can readily  produce large volumes of training pairs  of code + AST
using  compilers; however, the  code must  necessarily be  correct and
complete, to be compilable. We therefore have to artificially fragment
and noise-up this  code, to train our learned parser  to robustly deal with such
problems.  Our training approach resembles that of 
Pradel \etal~\citep{pradel2018deepbugs}. However, their goal was 
to \emph{detect} (not fix)  
specific types of bugs (e.g., accidentally swapped function arguments,
incorrect   binary  operators,  and   incorrect  operands   in  binary
operations) in
 \emph{otherwise syntactically correct  code};
our goal here is to leniently parse (thus, also repairing) 
and type syntactically malformed
code.

%must be syntactically correct; compilers are intolerant of syntax errors.
%For instance, every statement in Java must end with a semicolon (``;") and each open parenthesis (``(") must be accompanied with a close parenthesis (``)") (except within strings). Human beings are
%quite capable of figuring out code structure even with missing delimiters, and mentally ``correcting" for that, 
%but parsers are not. Constituency parsers for NL can be  trained for human-style ``lenience", 
%but we must incorporate such expectations into our training data. 
Second, the vocabulary
in code tends to be much larger~\citep{hellendoorn2017deep,karampatsis2020big} than natural language, thanks
to identifiers. Normally, larger vocabularies would present a challenge for learning to translate~\citep{hellendoorn2017deep}. 
However, for our purposes, identifiers fall into specific syntactic classes (variables, method names, type
names etc), and can be abstracted out into categories to simplify the parsing task, while keeping
vocabulary requirements modest, and brought back in later. 
%Though extensive vocabulary is a concerning issue for the code, the position of the tokens also carries useful information. %Therefore, it is possible to learn the syntactical rule for a programming language even with the vocabulary issue.
% Traditionally syntactical constituency parsing does not take care of the syntactical error or noise. Human is naturally much more tolerant of syntactical and even semantic diversity. Besides, unlike code, it is complicated to define syntactic rules for natural languages.
Finally, input code fragments (whether from \sofo or from student programs)
skew much longer than natural language sentences. 
As a result, syntax errors may arise from 
%However, there are errors with 
inter-related tokens hundreds or even thousands of tokens away from each other, \emph{e.g.;} one might forget the last closing curly brace (``\}") of a very long while loop. To handle the long-dependency problem, we use a two-stage pipeline, where both stages are trained to deal with  improper syntax. 
The first stage, \taseg, learns to identify and fix common patterns of block structuring in code. The output of this stage, is intended to clearly delineate the beginning and end of blocks, allowing easy segmenting of the code into statement-level fragments, which are typically 50-100 tokens in length. The next stage, \taparse, learns to parse statement-level fragments, fixing any simple syntax errors in the process.
%\vh{This feels duplicate/unnecessary here; preferably remove}
%The entire pipeline achieves state-of-the-art performance on the BlackBox dataset. 

\begin{figure}[htb]
\centering

  \centering
  \includegraphics[clip,trim=0cm 1cm 4cm 5.2cm, scale=0.35]{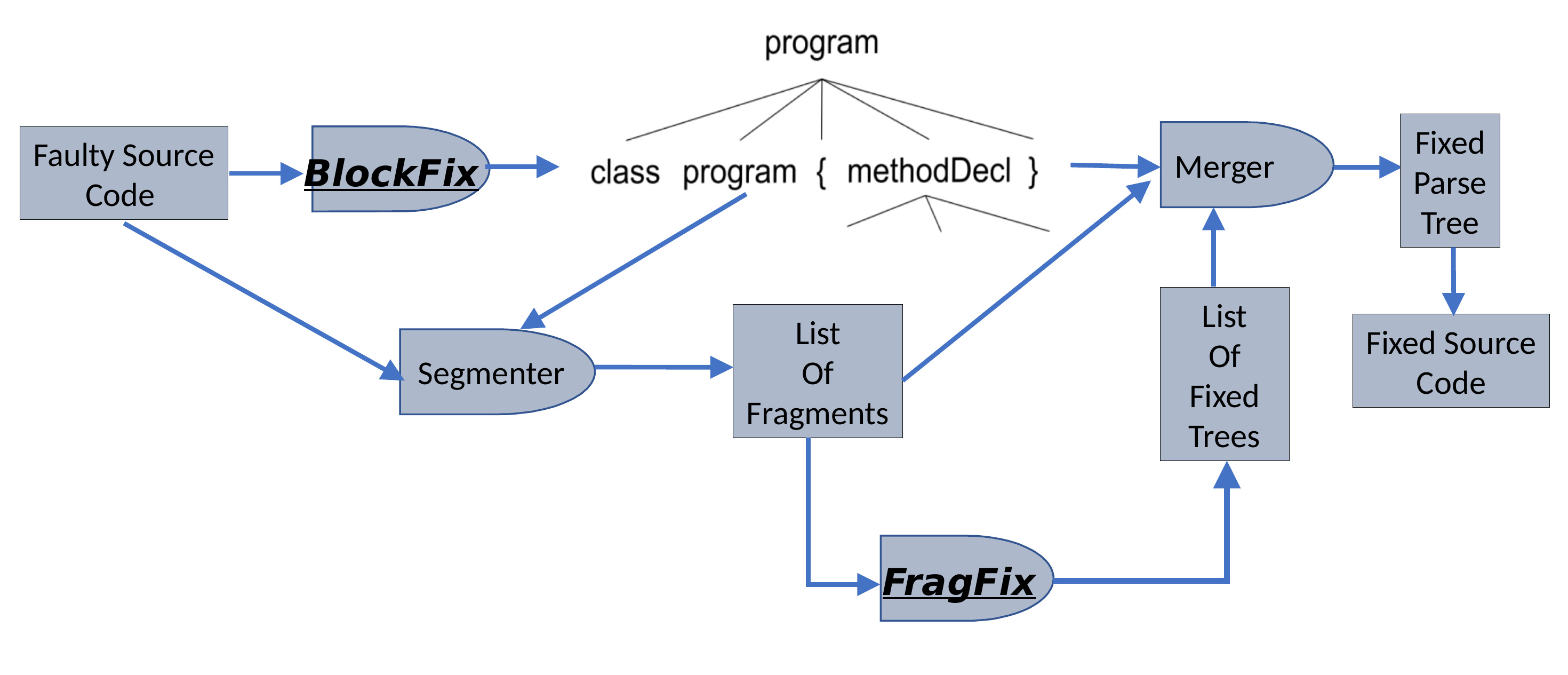}
  %\caption{Precision}
  %\label{fig:sub1}

\caption{\small  Full pipeline. {\underline {\bf \taseg}} and {\underline {\bf \taparse} }
are learned transformer models. For the \sofo parsing task we stop with
the ``Fixed Parse Tree''. For the Student code correction task we generate the fixed source code.}
\label{fig:overall}
\end{figure}

The details are in figure~\ref{fig:overall}. Given a (potentially faulty) code fragment, the learned \taseg model first identifies the proper nesting structure
of the blocks (details below) performing repairs as needed. Using this repaired nesting, the segmenter simply splits the code using block delimiters
("{\small\tt \{\}}") and statement delimiters ({\small\tt ";"}, and linespacing) to split the input code into fragments, while
retaining their point of origin within the block structure. The learned
\taparse model repairs and parses each fragment into a fixed tree. These fixed trees are then merged into the original block structure predicted by the the \taseg model. If repaired code is desired, the tree is ``unparsed" to the fixed code.

For the task of lenient typing (\emph{viz, } the \tatype model), 
we use the transformer architecture again, adapting the methods used in 
gradual typing applications of complete/correct fragments in Javascript~\citep{hellendoorn2018deep,  raychev2015predicting}. 
\tatype has a simpler task, requiring just a single-stage model as described below. 

%In the rest of this section, we begin with a brief introduction to the transformer model. After that we describe in sequence the training and operation of the \taseg, the \taparse and the \tatype models, all of which are transfomer-based. 

\subsection{Training Data}
%Although our goal is to perform lenient parsing on fragmentary, noisy student and \sofo code, and lenient typing on fragmented \sofo code,  
%our \emph{training data}  is not based on \sofo or student code; we wanted to avoid the challenging task 
%of creating labeled data  (ASTs and Types) from these fragmentary and noisy source code. 
Our training data consists of mostly clean, correct code from GitHub. 
We used code from 50 most popular projects from the 14,785-project dataset published by Allamanis and Sutton~\citep{githubCorpus2013}.\footnote{Allamanis and Sutton define popularity as the number of forks plus the number of watchers.}
Most of this code is professionally crafted,  with complete build environments. Thus it can be easily
compiled to produce ASTs and types. We use these to create training data. We also fragmented and 
added noise to this data, as described below, to make our parsing and typing models robust to fragmentation
and noise. Thus our approach relies on \emph{indirect supervision}, since the training
data originates from a different setting than the tasks. 

\subsection{Transfomers are all we need} 
\label{sec:model-parser}
%\subsubsection{Training Transformer Model for Lenient Parsing}  
The ``Transformer" is a DNN architecture originally developed for language modeling and translation. 
It relies exclusively on attention~\citep{vaswani2017attention} to model sequential dependency in language.  
Prior approaches to translation modeled this dependence using
recurrent   architectures,  \emph{viz.,} long short-term memory~\citep{hochreiter1997long}, and gated recurrent units~\citep{chung2014empirical}. %, and use an encoder-decoder mechanism for the task. 
These  recurrent  neural network (RNN) models  require ``back-propagation through time'' (BPTT) to recursively propagate
loss values over gradients, within the same recurrent units, so that long-distance dependencies could be captured
from the training set. While this approach is quite effective, the serial nature of BPTT
greatly
inhibits parallelism during training and use. RNNs also suffer from memory-loss when dealing with longer
dependencies. 
Bahdanau \textit{et al.} introduced \emph{attention}, wherein a fixed-length vector is used to identify and relate relevant part of the input to the target output~\citep{bahdanau2014neural}. This mechanism supplemented the recurrent structure and 
improved performance by enabling more direct calculation of dependencies. 

Standard, basic transformers essentially
take attention to the next level;
they outperform all older  models on various tasks.\footnote{See \url{https://ai.googleblog.com/2017/08/transformer-novel-neural-network.html}} 
They eliminate all recurrence, but multiply the attention mechanisms.  
We use transformers for all 3 of our tasks: lenient \sofo parsing and typing, and syntax error correction. 
As with older recurrent  (RNN) models, like RNNs, Transformers use input embedding layers to convert discrete 
sequential input tokens into sequences of continuous
vector embeddings, and softmax output layers to convert internal vector representations into output symbols. 
The output and input vocabularies are usually limited to control the input and output layer sizes. But the resemblance 
to RNN models ends here. 
Transformers primarily rely on layers of multi-headed attention, which can attend to multiple parts of the same sequence to help calculate a representation of the full sequence. These attention layers are interspersed with fully-connected feedforward layers that process the output from the attention layer. The non-linearities in the feedforward layer allow powerful and flexible combination of the elements upon which the incoming attention layer  is focused. 
Both the encoder and decoder part of the translation model use many such levels, each consisting of pairs of
stacked self-attention and fully-connected feedforward sub-levels. The encoder attends just to its own state (self-attention); the decoder attends both to any previously decoded symbols and to the encoded input at every layer in this stack.
This structure allow the model to attend to various parts of an input sequence, while
eschewing a recurrent architecture (and the attendant inherently non-parallel BPTT during training), which allows greatly increased model capacity, and more parallelism (and thus speed) during training. 
The number of sets of sub-layers in the encoder and decoder, as well as the number of heads is configurable depending on the task at hand. The original Vaswani \emph{et al} paper used stacks up 6 such sets, and 8 heads for attention in every layer. 
We use several different configurations for our various tasks, as described below; all are based on a configurable, open-source Transformer implementation freely available on GitHub.~\footnote{See \url{https://github.com/Lsdefine/attention-is-all-you-need-keras}}

\subsection{Recovering Block Structure: \taseg}
In programs with complicated nested blocks, 
code tokens can have very long-range syntax dependencies. 
%If the block structure is improperly nested, (\emph{e.g.,} with misplaced, extra
%or missing braces) errors arising from
%very long-range dependencies can result, and sometimes be difficult
%to spot and repair. 
The length of such dependencies can run into hundreds
of tokens; if there are errors,  
even very powerful DNN models can struggle to
identify and repair them. 
However, if the block structure were correct, it is possible to break code
into statement-level segments.

\taseg has the task of recovering the block structure: it learns to model common-block structuring
patterns, and repair nesting structure if necessary. The repaired structure allows
the code to be fragmented into forms that can be passed on to \taparse for repair, 
and then recombined into a whole AST. We illustrate with some examples. 
%; and the fragments
%themselves can be parsed using the lenient parser trained as above (\S~\ref{sec:lenient-sofo}). 
%\vh{Starting here, this feels duplicate/unnecessary; including next paragraph doesn't belong in the `BlockFix' section. Maybe make this the motivating example?} Finally, 
% thanks to the context-free, compositional
%nature of the grammar, 
%the ASTs fragments produced by \taparse can be re-constituted into a full AST of the
%original (unfragmented) code. % Along the way, errors noted in either stage can be repaired. 
%Using this two-stage (learned) approach, we can reach very high performance levels, both for \sofo fragment parsing, 
%and  for repairing
%student code. %(even \emph{without knowing the precise assignments}, as in SynFix~\cite{bhatia2016automated}). 

%are closely inter-related than the natural language and there are very strong long-dependency, still we can find pieces in code which are independent of another. 
%\prem{Is this an actual data from the BlueJ? Can you mention that?}

\begin{figure}[htb]
\captionsetup{aboveskip=-15pt,belowskip=0pt}
\centering
%\begin{subfigure}[t]{0.53\textwidth}
\begin{lstlisting}[language=Java,basicstyle=\scriptsize]
public class Batch{
    public static String subjects= "English,  Maths, Science" 
    public static void main(String args[]){
        if(subjects.length()>50) {
            System.out.println("subjects too long")
        }
    }
}
\end{lstlisting}
%\caption{Original}
%\label{jsorig}
%\end{subfigure}%
\hfill

\caption{Example Java program for segmentation.\vspace{-0.3cm}}
\label{fig:jexample}
\end{figure}

Consider the  code snippet in Figure~\ref{fig:jexample}.
Lines $2$ and $5$ are fragments that are syntactically independent of each other; 
despite missing the `;' ---our \taparse can fix and parse each separately. 
%We can process the ASTs of each statement first (line $2$ and $5$), 
%Pand processing of their ASTs does not depend on the processing of other parts of the code, \emph{e.g.,}: 
Now consider the fragment from line $4$-$6$. To generate the AST for this fragment, we can produce the ASTs for the statements of line $5$ (``{\small\tt System ... ;}") and ``if" clause (``{\small\tt if (...) \{ \}}'') separately. Now we know exactly where the AST of the former should go: between the two curly braces in the corresponding AST of the latter. If there were multiple statements, we would need to place the ASTs of all the statements sequentially in the block structure recovered by \taseg. 
%
%In this way, we can 1) \emph{fix and identify}
%the blocks code (with \taseg) 2) Use the block structure to fragment the code,  3) \emph{fix and parse} the fragments (with \taparse) 3) and 
% %Hence, we can 
%\emph{combine} the resulting fragment ASTs into the full AST.
% generate the AST for the complete fragment just by combining the smaller ASTs. 

%Similarly, we can generate AST of ``main" method defined in line number $3$ by inserting AST of the ``if-else" clause into the AST of the methods. 
The \taseg model requires
knowledge of typical block nesting structures, 
%long dependencies, 
and the discernment to fix errors therein.  
If there are unbalanced curly braces, this model repairs the code by inserting or removing braces
where they would be expected to appear. 
%and try to insert or remove braces to make the code correct. 
This is based on the assumption that the \emph{syntactic structures
commonly used in code are natural, with repeating patterns of nesting usage, which can be learned}. If we can learn
a model which knows these common patterns, it can also be trained to be ``forgiving" when curly braces are misused. 

%\prem{Confused by this: 
%Sometimes, even with balanced curly braces, it becomes very challenging to generate the fragments correctly,\emph{e.g.,}: consider line $2$ in Figure~\ref{fig:jexample}. The curly braces of those particular statements do not define any scope. Therefore, we should not generate fragments solely depending on the semicolon or braces. We use newlines to identify braces which define a scope of the program correctly. We split the program into lines and strip the default whitespace characters from the beginning and the end of the lines. The curly braces that define any scope should appear at the beginning or end of a line (there are very few exceptional cases though) if there were not written in a separate line. We identify those braces and fragment the program accordingly. However, if the program has a syntactical error, this approach can still fail (\emph{e.g.,} someone drops the semicolon from line 2). 
%
%On the contrary, if we drop semicolon from the statement which does not have curly braces do not affect the performance of the fragmentation. Dropping semicolon from line 5 may prevent us from separating the statements from the ``if" clause, but we may not face any problem because the number of the token is still less than 20 and the parser can process the full fragments at a time. Missing semicolon in consecutive statements merges those statements. However, the DNN base parser can generate the AST indicating the error if the length remains within a manageable range.
%}

\smallskip
\noindent\underline{\emph{Training \taseg}} %Model for Repairing Imbalanced Curly Braces}}
%We repair imbalanced curly braces by applying a transformer-based translation model.
%We train a separate transformer to learn the common patterns  of code nesting. This transformer
%model's attention window will include \emph{all} the tokens in th
%To build this model, we collected 1.5M random files from our  dataset. 
%For efficiency, we only used files with fewer than 500 tokens in the abstracted version for training. 
%Each of these files is parsed by the Eclipse JDT Parser, resulting in full ASTs. These
%pairs of files and full AST is our starting point to create training data. 

Since  \taseg has the sole task of modeling and repairing block nesting structure, 
it is trained with input-output pairs that just reflect this task. Consider the code below in figure~\ref{fig:jexample2}. 

\begin{figure}[h]
%\captionsetup{aboveskip=-15pt,belowskip=-5pt}
%\centering
%\begin{subfigure}[t]{0.53\textwidth}
\begin{lstlisting}[language=Java,basicstyle=\scriptsize]
public class TableListWriter extends HTMLListWriter{  
    public TableListWriter(File outputDir){ 
        super("Current Tables", "currenttables.html", "tables", outputDir); 
        if(ListClosing()) {
          WriteCloseMarkup()
          }
        else {
          CloseList();
        }
    }
}
\end{lstlisting}
%\caption{Original}
%\label{jsorig}
%\end{subfigure}%
%\hfill
\caption{Example Java program for abstraction.}
\label{fig:jexample2}
\end{figure}

\noindent We first abstract the input source code into an abstract form, like below: 

\smallskip

\noindent \fbox{\begin{minipage}{35.5em}

\noindent \footnotesize public class simple\_name extends simple\_name \{ public simple\_name paren\_expression \{ expression if paren\_expression \{ expression \} else \{ expression \} \} \}

\end{minipage}}

\smallskip

%\vh{This needs to be more concise. Maybe start with the example?} To train the segmenter, we created abstracted versions of the input file, and the output
%AST. 
%data from the same source we used for parsing and typing.
% However, instead of choosing data from the top 50 projects, we randomly chose 1M files from the dataset. 
%We abstract out both the input and the output of the segment
%to capture just the block-nesting structure. 
\noindent Almost everything is removed from the input, except
curly braces and keywords; we abstracted out all other identifiers, constants, expressions, and delimiters. 
%Consecutive sequences of expressions are collapsed into one abstract expression. 
%Identifiers are also abstracted. 
We can then simulate common structure-related syntax errors by corrupting this abstraction slightly and tasking the model with reproducing the original, uncorrupted (abstracted) code. Specifically, we add additional braces into half of the examples, while dropping some from the remaining half, split randomly between open and close curly braces for this noising step.\footnote{Anecdotally, additional braces are often next to existing braces; we therefore simulate this in 70\% of cases while inserting them in another random location for the rest.} Training on many such abstracted pairs allows the \taseg model to learn how syntactic constructs are most frequently nested in code. In all cases, the desired output shows where the mistakes are inserted, so the segmenter learns to be both lenient, and provide the correct fix if an error is detected.

The above process transforms a program with potential syntax errors into an abstracted structure with placeholders for all abstracted expressions. What happens to these bits that are removed when actually doing the task? These removed bits constitute the fragments that are sent along to \taparse for the next stage. 
Of course, we record and track where these fit into the abstracted input, so we can
reassemble the ASTs produced by \taparse from these code bits
into the abstract block-structure produced by \taseg.

Using this abstracted, noised segment training data, 
we train a transformer (translation) model.  This model is trained with files
whose abstracted versions are no more than 500 tokens; the dataset
includes around 1.5M such files. The 500-token limit fits inside the Transformer's
attention window, on a NVidia GeForce RTX 2080 Ti, without exceeding GPU memory; as explained below in
\autoref{sec:training-parser} this is adequate for our purposes. 
This model uses a
stack of two layers in both the encoder and the decoder, which we found was sufficient for our training setup.
Each layer includes multi-head self-attention and position-wise fully connected feed-forward layer, and in the case of  the decoder also multi-headed input (encoder) attention. Because of our chosen input/output abstractions, the input and output vocabulary size is limited to 54. As in the original Transformer architecture, all our layers use 512-dimensional states, which is split across 8 parallel heads for attention, and projected into 2048 dimensions (and back) in each pair of feed-forward layers.
Except for the number of layers $(N)$ (we use $N=2$ instead of $N=6$), we replicate all the hyper-parameters described in~\citep{vaswani2017attention}. We use an Adam optimizer as a learner with 4000 warm-up steps. We apply layer normalization after each sublayer. To prevent overfitting, we employ residual dropout ($0.1$) for regularization. We also add positional encoding to the inputs and vary the learning rate following the recommendation of Vaswani \etal~\citep{vaswani2017attention}. We trained our model for $10$ epochs with a batch size of $64$ fragments. 
 % \vh{64 what? Can't be tokens; there are way more in a file.}\toufique{64 fragments. Sorry for the confusion}
%we use for parsing task with all hyperparameter unchanged. 
The limited vocabulary allows most of the model capacity to be used for learning and repairing nesting structures, 
which helps the performance. 
% much simpler compared to natural language translation. 
%However, the abstraction can be difficult if there are syntactical errors in the code primarily error with the semicolon and parenthesis, e.g., if there is a missing semicolon in an expression, that should not be abstracted at first place and in the later phase it would yield the same number of tokens as it had initially. Since the model is trained to balance the curly braces, it still should perform reasonably well with a small sequence. 

Sometimes it is not clear exactly where a close curly should be added. Consider the example in Table~\ref{bfix} from student error code dataset. Should the close ``\}'' be added before, or after the canvas.draw()? Both would lead to correct parse. In this case, \taseg learns from the training set that nested loops are usually closed all at the same time, and proposes this fix. 
%Of course it could be other way, but based on the (large) training set, \taseg correctly proposes this fix. 
%
Both Eclipse and IntelliJ IDEs recommend adding the close ``\}'' after canvas.draw(), which is not the correct fix 
here. 

\begin{table}[]
\centering
\caption{\taseg fixing nesting error}
\label{bfix}
\begin{tabular}{|l|l|}
\hline
\multicolumn{1}{|c|}{Version with Error} & \multicolumn{1}{c|}{Actual Fixed Error} \\ \hline

\begin{lstlisting}[language=Java,basicstyle=\scriptsize]                                         
public class ImageBasics
{
  public static void main(String[] args)
  {    
    APImage cv = new APImage(255,255);
    for(int x = 0; x<255; x++)
    {
      for(int y = 0; y<255; y++)
      {
        Pixel p = cv.getPixel(x,y);
        p.setRed(x);
      }
    canvas.draw(); 
  }
}  
\end{lstlisting}

&

\begin{lstlisting}[language=Java,basicstyle=\scriptsize]                                         
public class ImageBasics
{
  public static void main(String[] args)
  {    
    APImage cv = new APImage(255,255);
    for(int x = 0; x<255; x++)
    {
      for(int y = 0; y<255; y++)
      {
        Pixel p = cv.getPixel(x,y);
        p.setRed(x);
      }
    }          
    canvas.draw(); 
  }
}  
\end{lstlisting}

\\ \hline
\end{tabular}
\end{table}

\label{sec:lenient-sofo}

\subsection{Recovering ASTs from fragments: \taparse}
\label{sec:training-parser}
The next step is to train a lenient \emph{fragment}  parser that can fix \& parse fragments. 
We first gathered all the Java files with fewer than 10,000 tokens and produced the ASTs using Eclipse JDT. 
This limit ensures that the abstracted versions of these files are small enough to be processed by \taseg; in
practice, this limit works well.  
Most Stackoverflow fragments are much smaller than 10,000 tokens, and  only 0.6\% of  the student programs
exceed this limit. 
%Since all the projects are from professional developers and very popular, the code is mostly syntax error-free.  
These (mostly) correct, well-structured programs are easily broken into
%From the code and AST pairs, we generated 
fragments using the semicolon (``;") and curly braces (``\{" and ``\}")
as breaking-points. We tried to keep the fragment-length roughly uniform, with limited variance, and removed
duplicates, % \prem{Check previous sentence!}
%We maintained a proportion between the number of tokens in each file and the number of fragments. 
%We also removed duplicated fragments in the same file. 
Our belief was that these fragments, despite their disparate origins % and the flexibility of programming languages, 
would nevertheless have repeating  syntactic patterns that \taparse would learn to capture, even out of context.  We collect 2M such fragments. 

Our next task was to train \taparse to be \emph{lenient} with respect to common errors. 
 Santos \textit{et al.}~\citep{santos2018syntax}  find that, in BlackBox, most (57.4\%) of syntax errors arise from
 single-token errors (extra, missing, or wrong tokens).
%  \vh{Just to make sure: insertions were needed to fix, or an insertion was the problem? If the latter, maybe choose a different word for ``deletions'' (omissions?)} 
Extra tokens accounted for about 23\% of single-token errors; 
 missing tokens for about 69\%, and substitutions for about 8\%.
 Based on a manual examination of a small sample (around 1,000 examples) of the data,
 we noted that the vast majority of these single-token errors centered around certain tokens we may call separators; tokens such as commas, semicolons, periods, all types of brackets (``{\small\tt [\{()\}]}''), and the string separator ``{\small\tt +}''. To gain a better understanding, we examined the errors from 200 randomly selected student programs. We found nine significant categories. The errors involving a specific category includes deletion, insertion or misplacement of that particular separator. Table~\ref{err} presents our findings. The other category includes cases with spaces between variable names, missing quotation marks, and some misplaced symbol which can not be covered by eight major classes. Of our 2M fragments, we mutated approximately half.  We sampled locations within fragments uniformly  at random
to inject these mutations.  In the majority of cases, each fragment gets one mutation because our primary goal is to solve a single token error. However, to make the model robust, we sometimes injected two mutations (~30\% of the mutated fragments have two mutations).

\begin{table}[b]
\centering
\caption{Categories of student code error}
\label{err}

\begin{tabular}{|L{3cm}|C{2.5cm}|C{2.5cm}|}
\hline
\multicolumn{1}{|c|}{\textbf{Category}} & \textbf{Total count} & \textbf{In percentage (\%)} \\ \hline
Semicolon                               & 73                   & 36.5                        \\ \hline
Curly brace                             & 39                   & 19.5                        \\ \hline
Parenthesis                             & 28                   & 14.0                        \\ \hline
Arithmetic operator                     & 20                   & 10.0                        \\ \hline
Keyword                                 & 9                    & 4.5                         \\ \hline
Comma                                   & 7                    & 3.5                         \\ \hline
Missing datatype                        & 6                    & 3.0                         \\ \hline
Bracket                                 & 2                    & 1.0                         \\ \hline
Others                                  & 16                   & 8.0                         \\ \hline
\end{tabular}
\end{table}

%\vh{The below could be motivated better. Maybe something like: We therefore decided to focus on [...] in this work, as learning to recover from these errors provides the most immediate path towards a time-saving ``lenient'' parser.}

Based on the commonality of these errors, 
%In this work, 
we sought to teach \taparse to robustly recover from them. 
 %mostly contained a single token errors: \prem{do  deleting the token, around 25\% error occurs due to insertion of the token, whereas the number of errors due to substitutions is negligible. 
 %We incorporated two significant types of error: 
%consisting of random 
%insertion and deletion of tokens. Since our focus is on syntactic lenience, 
%To help our lenient parser learn to be ``forgiving" of such errors 
We therefore inject occurrences of these errors into the input source code in our training data (details of error-injection below). 
These corrupted  inputs were paired with the original, correct AST, which indicated the location of the error, and
it's repair,  as illustrated below: 
%After that, we tried to reflect the change into the ASTs, \emph{e.g}: consider the following training insstance, consisting of a code fragment and desired output AST: \prem{We also introduced extra tokens? how about $\ldots$ dots}
 
\smallskip

\noindent \fbox{\begin{minipage}{35.5em}

\noindent{\scriptsize\bf Code:} {\scriptsize\tt int x = 0}

\scriptsize  {\bf AST:} (\#VariableDeclarationStatement (\#PrimitiveType ) (\#VariableDeclarationFragment (\#SimpleName)\\
 (\#PunctTerminal) (\#NumberLiteral)) \emph{\textbf{(\#missing-semicolon)}} )
\end{minipage}}
\smallskip

\noindent The ``redeemed" AST in this training sample clearly signals the {\small \em\bf \#missing-semicolon} in the code fragment, which can be used to repair the code. 
%Unlike past work which only corrects errors, without offering
%context, our learner is trained to produce a complete AST, with error indicated in context.
%\vh{It is not terribly clear to me why this is different: you still just show the faulty location, just like past work. Having an AST isn't necessarily helpful as a teaching device.}
%This not only enables automated error correction, but also provides a path towards
%a helpful explanation (which remains future work),
Also note that we drop concrete code tokens from the desired output, retaining just the AST nodes; 
this reduces not just the size of the output sequence (the serialized ``parse'') but also of the output vocabulary, simplifying the learning task, and allowing us to better leverage the capacity of the DNN learner.  
During actual parsing, we know the true input tokens, and can reproduce the full ASTs by inserting the tokens into the output in the same order.
%\vh{This might also better fit in the more general data discussion}. 
We also abstract all the numeric values to {\small\tt 0} and all strings and characters to their empty values ({\small\tt "", `'}), since these values tend to increase the vocabulary size without contributing to the structure of the AST. 

Regarding simulating typical errors, we observed from an examination of the data (both \sofo and student code), that erroneous inserts of separators do not occur uniformly at random locations; instead, they predominantly occur next to other separators. So, we often see ``stutter" errors of the form ``{\small\tt math.log(35.0))}''  
or "{\small\tt x = 0;\}\}}'', with repeated separators, but 
rarely ones of the form ``{\small\tt math.(log(35.0)}''  or  "{\small\tt x =\} 0;\}}'. To prioritize learning to ``forgive"  such errors, we prepared training data similarly biased towards stutters. 
To mimic these errors, we randomly choose separators within fragment as described above, and with 70\% probability repeat that separator, while the remaining 30\% are randomly inserted elsewhere in the code. These errors are paired with the ``redeemed" AST indicating the position of the extra
token. This data trains our parser to produce an AST that both indicates what was wrong, where, and how
to fix it.  

%The fundamental difference between the previous models and transformer is that because of the multi-head self-attention transformer can remove the recurrent computation of encoder and decoder and still receives more information from the different parts of the sequence. 
%In our experiment, we use a stack of 2 identical layers in both the encoder and the decoder. Each encoder layer consists of two sublayers, i.e., multi-head self-attention layer and position-wise fully connected feed-forward layer. We also have an embedding layer with 512 dimensions. The decoder is exactly like the encoder with one extra sublayer, i.e., multi-head attention layer for retrieving information from different part of the encoder output. 

The transformer architecture  for \taparse is identical to that used for \taseg; only the input and output configurations
are slightly different. Our output now includes just the vocabulary of possible AST nodes in the tree, and excludes all input tokens. For Java, this means that size of our output vocabulary is just $95$ tokens. That simplifies the translation task greatly; we found that transformer with $2$ layers is sufficient. 
Our encoder input does include regular code tokens, which can be highly diverse; thus, we create a limited input vocabulary of the $64,833$ most common tokens by discarding tokens which appear less than $12$ times in the training corpus. We use the same training regime here as for \taseg. 

\subsection{Final Lenient Parsing Pipeline}
We summarize the entire pipeline using the algorithm below, specifically for processing student code. The \sofo parsing is slightly different, and is explained afterwards.
For student code: first, we check if the input code fragment has balanced braces, using a simple counter-based algorithm (line 1,2). If not, (line 3) it is sent through \taseg which fixes the block structure.  
Next (5), any block delimiters, semi-colons, and linefeeds are used
as markers to identify locations where the input source code can be split into fragments. These markers are also used later to reassemble the fragments. Line 6 splits the input code using these markers into a list of fragments. Each fragment is then parsed (leniently) by \taparse (lines 8,9) and then the resulting fragments are used to re-assemble the full AST (line 10). Finally, using the indicated errors (missing/extra operators, delimiters etc), the repaired source code is generated in line 12. 

For the \sofo parsing task, there are two differences. First, many \sofo fragments are quite short. Since \taparse can manage fragments shorter than 40 tokens, we just skip the \taseg phase for these. Second, since we only need the AST, we skip the code generation step on line 12.

	{
 
		\begin{algorithm}[h]	
  \SetKwData{Left}{left}\SetKwData{This}{this}\SetKwData{Up}{up}
\SetKwFunction{Union}{Union}\SetKwFunction{FindCompress}{FindCompress}
\SetKwInOut{Input}{input}\SetKwInOut{Output}{output}
			\caption{Steps Followed for Student Code Correction}
 			\Input{Code fragment $P$}
			\Output{Fixed-up Code Fragment ${\mathcal P}$}
			        $abs \leftarrow FindBraces(P)$\;
			        \If{NotBalancedBraces(abs)} {
			        $abs \leftarrow \taseg(P)$\;
			        }
			      	$segs \leftarrow segment(abs, P)$\;
			      	$frags \leftarrow splitProgram(segs, P)$\;
				$AST \leftarrow initializeAST(abs)$\;
			      	\For{$frag \in frags$ $(in~order)$} {
						$fragAST \leftarrow \taparse(frag)$\;
						$AST \leftarrow Ins(fragAST, abs, segs, AST)$\; 		      	
						}				  
			${\mathcal P} \leftarrow GenerateCodeFrom(AST)$\;
			 \label{alg:pipeline}
		\end{algorithm}
 
	}

\subsection{Lenient Typing}
\label{sec:training-typer}
Many \sofo fragments omit declarations or {\tt import}s. Therefore, using even a
simple fragment is challenging, since identifier \emph{types}  cannot be easily derived.
% without this knowledge, 
%an IDE (for a statically typed language like Java) cannot assist in suggesting declarations, imports, etc.
%ometimes it becomes difficult to use them because of not knowing the exact type of the identifier for statically typed language like Java. 
Prior work~\citep{raychev2015predicting} showed that it is possible to guess and type annotations for gradually typed languages
such as Typescript. 
%introduced several approaches to automatically typed dynamically typed languages like JavaScript and TypeScript. %Raychev \textit{et al.} built a probabilistic learning model from big code to predict the types~\cite{raychev2015predicting}. 
Hellendoorn \emph{et al.}~\citep{hellendoorn2018deep,malik2019nl2type} use DNNs to predict types,  
formulating this task
as a sequence tagging problem because there is a one to one mapping between the input token and types~\citep{hellendoorn2018deep}, They used an RNN architecture. 
with non-identifiers receiving an empty annotation. 
None of these approaches have been applied to Java \sofo fragments that lack
imports and declarations, yet having type information for a fragment may enable a downstream IDE to 
suggest declarations, imports \emph{etc} (or even renamings for variables in the fragment to variables
of the same type that are available, and in scope) when re-using that fragment.

We followed an approach similar to~\citep{hellendoorn2018deep}, except using the trans\-former-based model instead of an RNN. Our training data consists of the same projects as before; we used Eclipse JDT to derive the types for all identifiers in these Java files, while marking non-identifiers (\emph{e.g.,} keywords, operators, delimiters) with a special `no-type' symbol (in the following example, we use a special symbol ``$\sim$''). After generating types for every token (in all complete files), we created random (cross-project) fragments for training data, as we did for the parsing task, with corresponding types as derived by JDT. In total, we extract ca. 2 million fragments from the projects with the desired types, all similar to this pair below:

\smallskip
\noindent \fbox{\begin{minipage}{35.5em}
\noindent{\footnotesize\tt if ( \emph{\textbf{something}} ) \{ Object \emph{\textbf{o}} = new Object ( ) ; \}}

\footnotesize $\sim$ $\sim$ boolean $\sim$ $\sim$ $\sim$ java.lang.Object $\sim$ $\sim$ $\sim$ $\sim$ $\sim$ $\sim$ $\sim$
\end{minipage}}
\smallskip

%\vh{I'd prefer to just annotate no-type tokens with -- (an emdash, preferably); it's easier for new-comers than `O'}
Out of $14$ tokens in this fragment, $2$ are identifiers for which types are provided; one primitive and one fully quantified. The other tokens are deterministically tagged with ``$\sim$" to simplify the model's task.

%Unlike parsing, we did not replace all the numeric values with $0$ this time because numeric values carry essential information about their type 
%\prem{Seems weird. Why not use one canonical
%value for each type, viz., 0, 0.0, "", 0b, etc?}
%\toufique{I did not do it because it did not create any technical challenge here and seems like it was not required also. I have not explained well I guess. During Parsing 0, 0.0, 0x56 everything falls under one class--"number literal". During Typing its not the case 0 can be integer, long, short even big decimal. The type information flows from the declation and even from the position of the token e.g. parameter of a function which takes float as input. On the back of my mind, one this was going on that may not be true. I feel like by doing it I add biasness to particular token. That means most of the 0s will be integer and whener the model see a zero it may assign the type int to it. Another reason of not doing it we have very high accuracy on these primitive types. On the otherhand, we we don't abstract the string, we will face technical challenges due to one to one mapping.}
% However, we followed the same rule for string contents like parsing. 

\smallskip             
\noindent{\emph{Training Transformer Model for Lenient Typing:}}
As before, we used a transformer-based model for typing of Java  fragments from \sofo. However, we formulate
the typing task as a sequence-tagging problem (similar to part-of-speech tagging, Named Entity Recognition \emph{etc.} ) since the input and output lengths are always identical, unlike with the translation task. Also, the output
vocabulary (the set of possible types) is much larger. 
Therefore, the translation model used for parsing is not directly applicable. 
Sequence tagging is in some ways an easier task than translation: we do not need to digest the full input sequence.  
Types can mostly be assigned based on local information, so there is no need for a full encoder mechanism to
encode the full input; the task can be performed with a single ``decoder'' element. 
%\prem{Input and ouput vocabulary counts!~!!}\toufique{mentioned a few lines later}
%in the encoder and then generate output token in the decoder. Sequence tagging is much more memory efficient than translator because it does not require an encoder-decoder mechanism. Therefore, we can process more token at a time than the translation model.
% If we only use the decoder from the translation model, it can be used as a sequence tagger. 
In the absence of the encoder element, the decoder simply attends (using multiple heads) to various tokens
of the input sequence, as it generates tags (types) on the output. 
%pNote that the fundamental difference between these two models is in the attention layer. Since the encoder is missing in this architecture, only multi-head self-attention is applied to the input sequence, and there is no multi-head attention for retrieving information from the encoder output.   
%T

For this task, the hyper-parameters are set as recommended by Vaswani \textit{et al.}~\citep{vaswani2017attention}. 
For the single ``decoder" element, 
we use $6$ layers (each consisting of  multi-head attention + feedforward) 
instead of $2$ to provide enough capacity to model the much larger input and output vocabulary. We keep all other hyperparameters unchanged except for the learning rate and warm-up steps. We set the initial learning rate as $0.2$ and warm-up steps at $1000$. We also use a warm restart for the learning rate~\citep{loshchilov2016sgdr} by resetting the learning rate to its initial value after each epoch. Note that because of the one-to-one mapping, the length of the input and output sequence must be same. We include a token into the input vocabulary if the token appears at least $35$ times in the training corpus; the cutoff value for output (type) vocabulary is $50$, making the size of the input and output vocabulary $40,316$ and $18,673$ respectively. We prevent gradient updating for the non-type token to simplify the learning process. We trained the model for 10 epochs with batch size of 4000 tokens.

\section{Evaluation \& Results}
\label{eval}
%--evaluation approach and datasets. 
%--- for parsing 
%--for typing
%---for error correction.

We used a mixed methods approach to evaluating our 3 tasks: \sofo parsing, student code correction, and
\sofo typing, based on the characteristics of each task. Table~\ref{ep} presents the detailed evaluation plan of our research.

\begin{table}[h]
\centering
\caption{Detailed evaluation plan}
\label{ep}
\begin{tabular}{|L{2cm}|L{2cm}|L{2.25cm}|L{2.75cm}|L{2.5cm}|}
\hline
\multicolumn{1}{|c|}{\textbf{Tasks}}   &\multicolumn{1}{c|}{\textbf{Models}}                   & \multicolumn{1}{c|}{\textbf{Dataset}} & \multicolumn{1}{c|}{\textbf{Baseline}} & \multicolumn{1}{c|}{\textbf{Evaluation Approach}} \\ \hline
\multirow{3}{*}{Lenient Parsing}       & \taparse                           & Synthetic                             & N/A                                    & Quantative                                        \\ \cline{2-5} 
                                       & \taseg                          & Synthetic                             & N/A                                    & Quantative                                        \\ \cline{2-5} 
                                       & \taseg + \taparse                  & \sofo                                  & Wrap \& Parse                          & Quantative-manual                                 \\ \hline
\multirow{2}{*}{Lenient Typing}        & \multirow{2}{*}{\tatype}          & Synthetic                             & N/A                                    & Quantative                                        \\ \cline{3-5} 
                                       &                                   & \sofo & N/A                     & Quantative-manual                                 \\ \hline
Student Code Fixing & \taseg + \taparse & Blackbox~\citep{brown2014blackbox}              & DeepFix~\citep{gupta2017deepfix}                                & Quantative                                        \\ \cline{4-5} 
                                       &                                   &                                       & Santos \etal~\citep{santos2018syntax}            & Quantative (reported from paper)                  \\ \hline
\end{tabular}
\end{table}

First, to evaluate lenient parsing of \sofo fragments, we used a combination of automated and manual methods. We have two different datasets for the evaluation. 
For \taseg and \taparse, we use a synthetic dataset, which includes mutated  (and untouched)
fragments from GitHub; the ``golden" parsed AST is produced by Eclipse, and is compared
to our lenient parser. Note here, however, that the lenient parser's task is to process mutated \emph{and} untouched fragments to yield an AST; this AST
is being compared against the AST produced by Eclipse from the \emph{non-mutated}, original, code. In this way, we evaluate the power of the
lenient parse components  to parse both untouched and synthetically mutated code fragments. 
We note here that \taseg and \taparse phases are unique to our approach, 
and do not have any comparable baselines. 

Second, for parsing the \sofo fragments, 
we used simple \emph{wrap \& parse} trick on \sofo fragments to baseline our model: wrap the code in a function skeleton, and process with Eclipse JDT parser. 
If the AST generated by our lenient parser matches precisely to the AST generated by wrap \& parse, we considered them correct. 
For the \sofo fragments on which the wrap \& parse trick failed, we had to check by hand. 
 %had to manually validate them because we do not have any ground truth for those fragments. 
 %However, we did not apply wrap \& parse to the synthetic dataset. Since we generated the fragments with Eclipse JDT, we already verified the synthetic fragments with JDT before including them to the \taparse training and held-out set.
%  Wrap \& parse is nothing but a simple way to check whether we can parse the fragments with Eclipse JDT. Therefore, we already know that if the fragment is mutated, it can not be parsed by Eclipsed JDT, and unmutated one can be parsed with the correct wrapper. 
%  There is another model \taseg that helps to repair the nesting structure, and it does not produce any AST. To the best of our knowledge, there is no such other model. Therefore, we do not have any baseline for this particular task.         

Third,  for the lenient typing task in \sofo, we used only a manual evaluation. We are not aware of any previous approach for finding the types in fragments of code, 
that would be suitable as a baseline. We have a synthetic dataset also for the evaluation of the model, as with parsing: the lenient typer is asked to type
mutated GitHub fragments which are evaluated against types produced by the Eclipse JDT from the original, un-mutated code. 

Finally, for the student code correction task, we used a fully automated evaluation. We explicitly compare our tool with DeepFix~\citep{gupta2017deepfix} and the tool proposed by Santos \etal~\citep{santos2018syntax}. The rationale and results are presented separately in sections below.

\subsection{Performance of Lenient Parsing} 
\label{sec:sofoparse}
As clarified in Section~\ref{sec:training-parser}, the lenient parser comprises two models (i.e.,     \taparse and \taseg). \taparse was trained with 2M fragments collected from GitHub. After training, we first evaluate it on 50,000 test fragments, consisting of about 29,300 mutated and 20,700 untouched fragments. 
from GitHub. These test fragments  originated from correct code that could be parsed automatically by Eclipse JDT to create ASTs as a ``golden" benchmark. 
We ensured that the test data didn't have any fragments duplicated from the training data. 
With NVidia GeForce RTX 2080 Ti GPU, we could able to train one epoch in about two hours using the transformer-based model. For \emph{mutated} 
held-out fragments, we found that 94.5\% of the outputs from the trained lenient parser accurately (exactly) matched the output from Eclipse JDT.
For the untouched fragments, we achieved an accuracy of 95.7\%. This supports the claim that the lenient parser is able to overcome
fragmentation \& mutation. 
We also observed that \taparse's accuracy decreased with fragment length, with performance sharply decreasing above a 40-token length.
Of course, our main interest is the performance on actual \sofo fragments, which may be syntactically erroneous, and thus impossible to parse directly with Eclipse JDT; our approach to this is described next.
To evaluate our \taseg model, we also used examples from GitHub. In this case, however, since segmenting
correct code is trivial, we only evaluated on incorrect code. 
 We trained \taseg with 1.5M training examples and achieved around 76\% accuracy on our test set, again using a transformer-based model.

\smallskip             
\noindent{\underline{\emph{Parsing performance on \sofo}}}
A key goal of lenient parsing is to correctly process 
malformed \sofo fragments.  
These, by definition, could not be automatically parsed
and so required manual checking. 
%Given the manual effort required for evaluating results over erroneous fragments
Our evaluation of \sofo test samples is limited by required human effort. Still, we sought a
a sufficiently large \& representative sample to get a good estimate of 
the performance. 
We collected the \sofo fragments from the public Google BigQuery 
dataset.~\footnote{See \url{https://cloud.google.com/bigquery/public-data/}}
For this experiment, we collected the answers for questions tagged with ``Java.'' After that, we isolated the fragments using ``$<$code$>$'' tag used for presenting code snippet. We randomly chose
a total of 200 fragments with various lengths for evaluation. 
%In general, \sofo fragments 
%are slightly different from our training and test data both in nature and length. 
%Though most of the \sofo fragments are short, still there are a significant amount of very long fragments, and their length is not uniformly distributed. We mutated the data to incorporate the syntactical errors, but it is not possible to exhaust the list of never-ending syntactical errors the programmer generally make. Therefore, some syntactical errors may not present in the training set. 

Our goal here is to measure how often the lenient parser produces an AST that could easily be used by downstream tools, such as IDEs. For this reason, we believe the standard BLEU-score measure used for translation-based tools is unsuitable. Instead we used a repeatable, objective, 4 class categorization of outputs: {\bf a)} \emph{Correct}: the output AST exactly matched the correct AST. 
{\bf b)} \emph{Autofixed}: the model's output matched the correct AST after an automatic
 post-processing step of adding or removing close parens, ')', at the \emph{very end} of the output to balance all open '(' parens. \emph{No other change is allowed}. 
{\bf c)} \emph{Partial}: the output AST only matched
the top-level  node of the correct AST, and {\bf d)} \emph{Incorrect}: none of the above. The \emph{Correct} and
\emph{Autofixed} classes are designed to capture  cases which allow easy, automated downstream IDE use. 

One additional caveat: in the absence of context, it's virtually impossible to distinguish between field (class member)
declarations and variable (local variable) declarations in small fragments. When pasting in a parsed AST fragment, it should be quite 
possible for an IDE to adapt the declaration form as needed; so in our evaluations, we ignored this
distinction. Either was considered correct. Consider the following fragment. Though it's a field declaration, \taparse predicts it as a variable declaration.

\begin{smcodetabbing}
List <Class<?>> defaultGroupSequence = new ArrayList <Class<?>> ( ) ;
\end{smcodetabbing}

Given a \sofo fragment, we used a two-stage scheme for checking ASTs produced by the lenient parser. First, we
attempted to embed the fragment within a  class ({\small\tt class \emph{classname} \{ \ldots \}})  or method 
({\small\tt void \emph{methodname} () \{ \ldots \}}) wrapper, thus turning it into a unit potentially parseable by JDT. If
the JDT would parse the fragment within such a wrapping, we had the exact AST for the fragment, and use
that as the correct baseline. If such a wrapper could not be found, we manually evaluated the lenient parser output. 
Of the 200 fragments, 123 could be parsed after wrapping by JDT, and 55 could not. The remaining 22 fragments were
not Java,  but XML, Gradle, data \emph{etc}. The outputs from the 178 Java fragments were categorized
as above; the \emph{correct} category was checked automatically whenever we had ``Golden" results from JDT. 
The rest  were manually checked
by the two authors independently, strictly following the protocol laid out above.

\begin{table}[]
\centering
\caption{\sofo fragments parseable by our approach but not by JDT}
\label{parse}
\begin{tabular}{|L{4.6cm}|L{4.6cm}|L{2.6cm}|}
\hline
\multicolumn{1}{|c|}{\textbf{Code Fragments}} & \multicolumn{1}{c|}{\textbf{By \taseg + \taparse}}                               & \multicolumn{1}{c|}{\textbf{Comments}}                  \\ \hline
 
\captionsetup{aboveskip=0pt,belowskip=0pt}
\centering

 \begin{lstlisting}[language=Java,basicstyle=\scriptsize\tt]
Optional<String> getIfExists() {
   ...  
   return Optional.empty();
}
\end{lstlisting}& 

\vspace{-2cm} 

( MethodDeclaration ... ( Block ... ExtraPunctuation(s) * ( ReturnStatement ... )... ) )

&

\vspace{-2.7cm} 

Unwanted ellipses in the fragment                                    \\ \hline
   
\begin{lstlisting}[language=Java,basicstyle=\scriptsize\tt]
case 3:
  if (price > 75 ) {
  totalPrice = price;
  } else {
  totalprice = 5.95 + price;
  }
  break; 
\end{lstlisting}

                                              & 
                                              
\vspace{-2.8cm}

                                              ( SwitchCase ... ( IfStatement ... ( InfixExpression ... ( Block ... ) ( BreakStatement ) 
                                              
                                              &
\vspace{-3.5cm}

                                               Missing initial part of switch statement                             \\ \hline

\begin{lstlisting}[language=Java,basicstyle=\scriptsize\tt]
str = str.replaceAll("0+$", "")
\end{lstlisting}

                                              &
                                              
\vspace{-1.8cm}                                               
                                              
                                               ( ExpressionStatement ... ( MissingSemicolon ) )                                   &

\vspace{-1.5cm}                                                
                                               
                                               Syntactical error because of missing semicolon                       \\ \hline

\begin{lstlisting}[language=Java,basicstyle=\scriptsize\tt]
getClass().getClassLoader().getResource("/resources")))
\end{lstlisting}

                                              &
                                              
\vspace{-1.8cm}                                               
                                              
                                               ( ExpressionStatemen ... ( ExtraPuntuation(s) * ... ( MissingSemicolon ) )               &

\vspace{-1.5cm}                                                
                                               
                                               Syntactical error because of missing semicolon and extra parentheses \\ \hline
                                             
\begin{lstlisting}[language=Java,basicstyle=\scriptsize\tt]
public BankAccount(double b, String n)
{
double balance = b;
\end{lstlisting}

                                              &
                                              
\vspace{-2cm}

                                               ( MethodDeclaration ...( MissingCloseCurly ) )                                       
                                               
                                               &
                                               
\vspace{-2.4cm}                                                
                                               
                                                Incompleteness of the block                                          \\ \hline
\end{tabular}
\end{table}

Of the 123 JDT-parseable
fragments, the lenient parser got 90 \emph{correct}s,  no \emph{autofixed}s, 27 \emph{partial}s, and
6 \emph{wrong}. Overall, the lenient parser, by itself, could produce ASTs in 126/178  cases
(or roughly 71\% of cases) in a form that was easily usable by downstream tools. This may seem like only a slight improvement on the 123 of the simple approach of wrapping and parsing with JDT, but the models did not actually solve the same fragments. Instead, on the 55 fragments on which JDT wouldn't work, the
lenient parser yielded 30 \emph{correct}, 6 \emph{autofix}, 16 \emph{partial}, and only 3 \emph{wrong} ASTs. In other words, while simple wrapping and then parsing works in about 69.1\% (123/178) of  cases, fragments that resist parsing with this trick can then be fed to our approach; this  \emph{\underline{fully automated}}
hybrid approach allows for parsing a total of 89.3\%, (123+36 = 159/178) of fragments in our sample (Wald confidence interval 85-94\%). A binomial test of difference of proportions yields
a p-value $<$ 0.00001 (n=178, ``heads" = 159 \emph{vs.} 123) for the null hypothesis
that the observed difference (between the combination approach and  simple wrapping+JDT parsing)
is due to random sampling error. Table~\ref{parse} presents some \sofo fragments along with the ASTs that are parseable by \taseg + \taparse but not by Eclipse JDT. Note that we present a concise form of the ASTs in the table. The complete ASTs are similar to the example presented in Section~\ref{sec:training-parser}.

\subsection{Performance of Lenient Typer}

%We evaluated the lenient typer using a similar method to the above. 
%
%\smallskip             
%\noindent{\underline{\emph{Performance of Lenient Typer on Test Set}}}
Our lenient typer was trained on about 2M training instances (49M tokens). To first get a sense of
the performance potential, we turned again to our 82K held-out fragments from the same data source, with their ``golden'' types from the JDT. We achieved 95\% top-1 and 99\% top-5 accuracy using transformer-based approach. For the top 1,000 most frequently-used types in our data, the top-1 and the top-5 accuracy are 97\% and 100\%; for primitive types in particular, \tatype is virtually infallible in this automatically created dataset. This makes sense given that Java is a statically typed language and these files contain no syntax errors; it implies that our model has accurately learned the distribution of types given tokens. The real test will be the actual \sofo fragments, where we need to manually check the predicted types.

\smallskip             
\noindent{\underline{\emph{Typing Performance on \sofo }}}
As before, we collected \sofo fragments  from 
the public Google BigQuery 
dataset and processed them using our learned lenient typer. The outputs in this
case, however, have to be checked entirely by hand, since most fragments lack the necessary build environment information (\emph{e.g.,} {\footnotesize\tt CLASSPATH, import}s) and cannot be automatically processed to get ``Golden types". We therefore selected 75 code fragments from highly rated answers (1000-3500 net positive votes). To get a broader diversity of samples, we collected these from 3 categories (25 from each): {\bf a)} \emph{Popular types} consisting of the 5 most popular (as identified by Qiu \emph{et al} ~\citep{qiu2016understanding}) Java classes:

\begin{enumerate}
	\item {\footnotesize\tt java.lang.String}
	\item {\footnotesize\tt java.lang.Override}
	\item {\footnotesize\tt java.util.List}
	\item {\footnotesize\tt java.lang.Exception}
	\item {\footnotesize\tt java.lang.Object}
\end{enumerate}

 %({\footnotesize\tt java.lang.String, java.lang.Override, java.util.List, java.lang.Exception, and java.lang.Object}), 
 
{\bf b)} \emph{Core Java types} consisting of any fragments tagged with only ``Java'' in \sofo, and {\bf c)} \emph{Android types} consisting of types that occur in the Android API, that don't fall to the other two categories. The Android category, in particular, can inform how the amount of available data affects the performance of our tool; Android API classes (though clearly important) were found in only 12 projects in our dataset, which accounted for about 4.5M tokens out of a total of corpus size of 52M tokens in all the projects. Therefore \tatype has a more limited exposure to Android API types during training. 
The other two categories were  well represented.

\begin{figure}[htb]
\centering

  \centering
  \includegraphics[scale=.30]{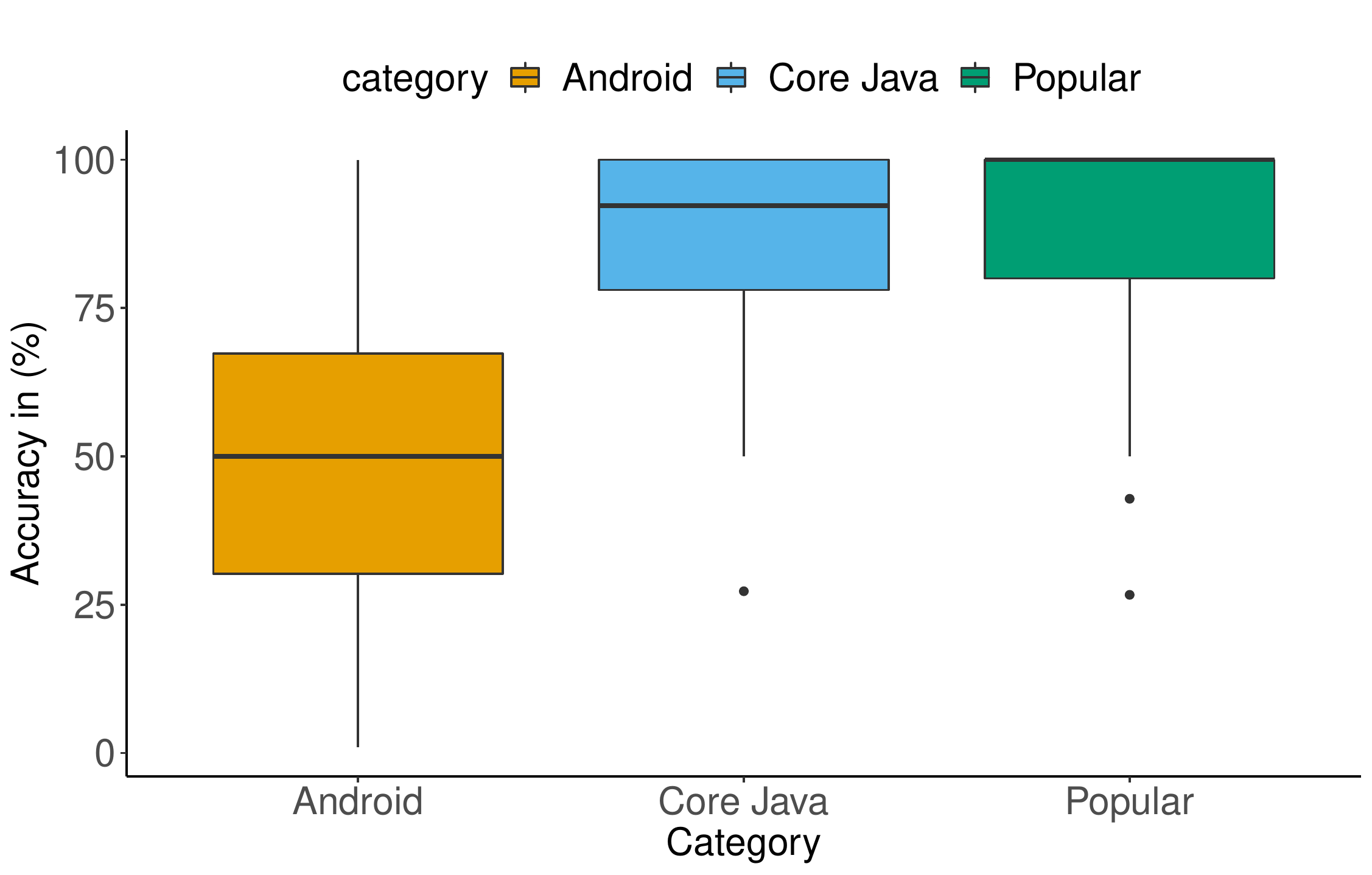}
  %\caption{Precision}
  %\label{fig:sub1}

\caption{Performance of DNN typer on different categories of \sofo fragments}
\label{javatype}
\end{figure}

We report our evaluation based on the proportion of identifiers in each fragment that were correctly 
typed. If used downstream in an IDE,
the incorrect identifiers would have to be fixed manually. 
This number is shown on the y-axis of Fig~\ref{javatype}
as a percentage. 
If all the identifiers in a fragment were labeled
correctly, the sample would score at 100\%. We break the scores into 3 groups by Category, and show
a boxplot for each. 
As can be seen, there appears to be a correlation between the amount of training data and performance.
We see the best performance for the \emph{Popular} category (median 100\%) and Core 
(median 90\%),  and a lower median for Android, around 50\%.
These results suggest that training \tatype on even larger datasets could further improve performance; we discuss
additional approaches to improve performance later (\S~\ref{sec:discussion}).

\subsection{Evaluation of Student Code Fixing}
\label{sec:studentcode}
%In this section, we briefly discuss the dataset and performance of our approach to fixing student code.  
%\smallskip             
%\noindent{\underline{\emph{Dataset for Fixing Student Code}}}
To evaluate the performance of the repair tool, we need realistic student programs with 
syntax errors, along with human-produced fixed versions. 
%To fairly compare our work, we wanted to apply our approach to the same 
We used the dataset used from the Blackbox~\citep{brown2014blackbox} repository, as used in Santos \textit{et al.}~\citep{santos2018syntax}; their work makes for a good baseline because this dataset is both very large and diverse.  
This Blackbox repository collects students' coding activity directly from the BlueJ Java IDE, which is designed to help beginners learn Java in programming classes~\citep{kolling2003bluej}. After obtaining permissions, we evaluated
our tool on this dataset.
%We contacted the admin and applied for access to the data of Blackbox. We also contacted the authors of~\cite{santos2018syntax}. On approval of the admins, Santos \textit{et al.}~\cite{santos2018syntax} shared their data with us. 
The dataset contains about 1.7M pairs of incorrect versions and the corresponding fixes. 
Santos \textit{et al.} found that many syntax errors (57\%, or \emph{circa} 1M pairs) can be corrected by a single token edit; they report their tool's performance only on files with just a single token error. 
They use the mean reciprocal rank (MRR) as a performance metric, which tracks the average of the inverse (reciprocal) of the rank of the correct solution found in an ordered list of solutions (i.e., repairs). We report both MRR and top-1 accuracy, for comparison.%  with them. 

Our general pipline for fixing student programs was described on page ~\pageref{alg:pipeline}.
%To fix the student program, we apply similar steps that we followed to generate the ASTs. 
%We first check if the erroneous student code fragment has balanced curly braces. If not, we process
%the student code with the \taseg, which abstracts out and repairs the nesting structure of the code. 
%We use a two-stage pipeline to correct student program. Whenever we process a file for code correction, we check whether the code has an equal number of open and close curly braces. If there are unbalanced curly braces, we send the code through the \taseg. The \taseg repairs the long dependencies by balancing the curly braces. 
%Then, we generate segments from the program, and process each with \taparse
%\prem{Need to revisit this after restructuring section 3} 
If the final \taparse stage signals missing/extra tokens, we use it to 
sample upto 5 most likely ASTs, to measure the MRR of the
correct answer. For each AST, we make the appropriate repairs, and check if any of those exactly match the correct
repair given in the dataset. If so, the reciprocal of the rank of the correct answer is recorded; otherwise,
we record zero. We also record the proportion of top-1 correct answers.

\smallskip             
\noindent{\underline{\emph{Performance of Student Code Fixer on Blackbox Dataset}}}
To get a performance estimate with tight bounds, we chose a very large random sample of 
 200,000 files out of the \emph{circa} 1M files with a single syntactical error. Our models achieve a 56.4\% top-1 accuracy rate and 0.58 MRR for the true fixes 
 %\vh{Those numbers are a bit odd; they suggest that there is virtually no gain after top-1. Maybe double-check}\toufique{about 3\% (600/20000) got repaired by alternative models. It mostly related to parenthesis related error. Suppose the model suggests removing extra paren but the programmer added one to make that work},
which is substantially higher than the reported MRR (0.46) by Santos \textit{et al.}~\citep{santos2018syntax}. 
Most of the time, the correct fix appears at the first rank; if not, we rarely see the fix in the top 5. 

We also applied our approach to files with 2 and 3 syntactical errors. Out of the remaining 700,000 pairs,
there are approximately 248K examples
with two syntax errors~\citep{santos2018syntax}, and
94K  files with 3 errors.  To estimate performance in these two categories, we chose 
50K files with 2 errors, and 50K with 3 errors.  In the two-error category, we measured 19\% top-1 accuracy, and
for the 3-error files, we noted 9\% top-1 accuracy. We note that Santos \emph{et al} do not consider files
with more than 1 error. 

In summary, out of all 1.7M programs with syntax errors, 57.4\% are single token errors
(as per Santos \emph{et al}~\citep{santos2018syntax}, Table 1), of which we can fix 56.4\% perfectly (top-1 correction), yielding an estimated top-1 fix rate for all files with syntax errors in the Blackbox of about 32\%.
If we consider the top-1 accuracy for up 3 syntactical errors, we estimate (using Santos' Table 1 estimates of 
proportions) that we could fix approximately 35\% of these files. In the remaining part of this section, we discuss various aspects of our model's performance.

\smallskip             
\noindent{\underline{\emph{Ablation: The \taseg's role}}}
%We applied a two-stage pipeline to fix the student program. In the first phase, we use a \taseg if required. However, using a DNN model without justification is not a wise decision because it is time-consuming and memory inefficient. 
We used \taseg to help fragment the code, since all DNNs (LSTMs or Transformers) struggle
with long-range syntax
dependencies. So how much does it actually help? 
We used 20,000 randomly chosen files to measure this effect.  We found quite a large number, 
4,925 (24.62\%) of files with unbalanced curly braces. Of these, our complete pipeline could fix 
3,253 (66.25\%) cases, yet \taparse \emph{per se}, without \taseg, could
only fix 36 (0.9\%)! For the remaining 15,075 files, \taparse did still work fairly well, incurring an overall MRR drop from 0.58 to 0.42. Thus, we believe that \taseg plays a useful and complementary role.

\begin{figure}[htb]
\centering

  \centering
  \includegraphics[scale=.32]{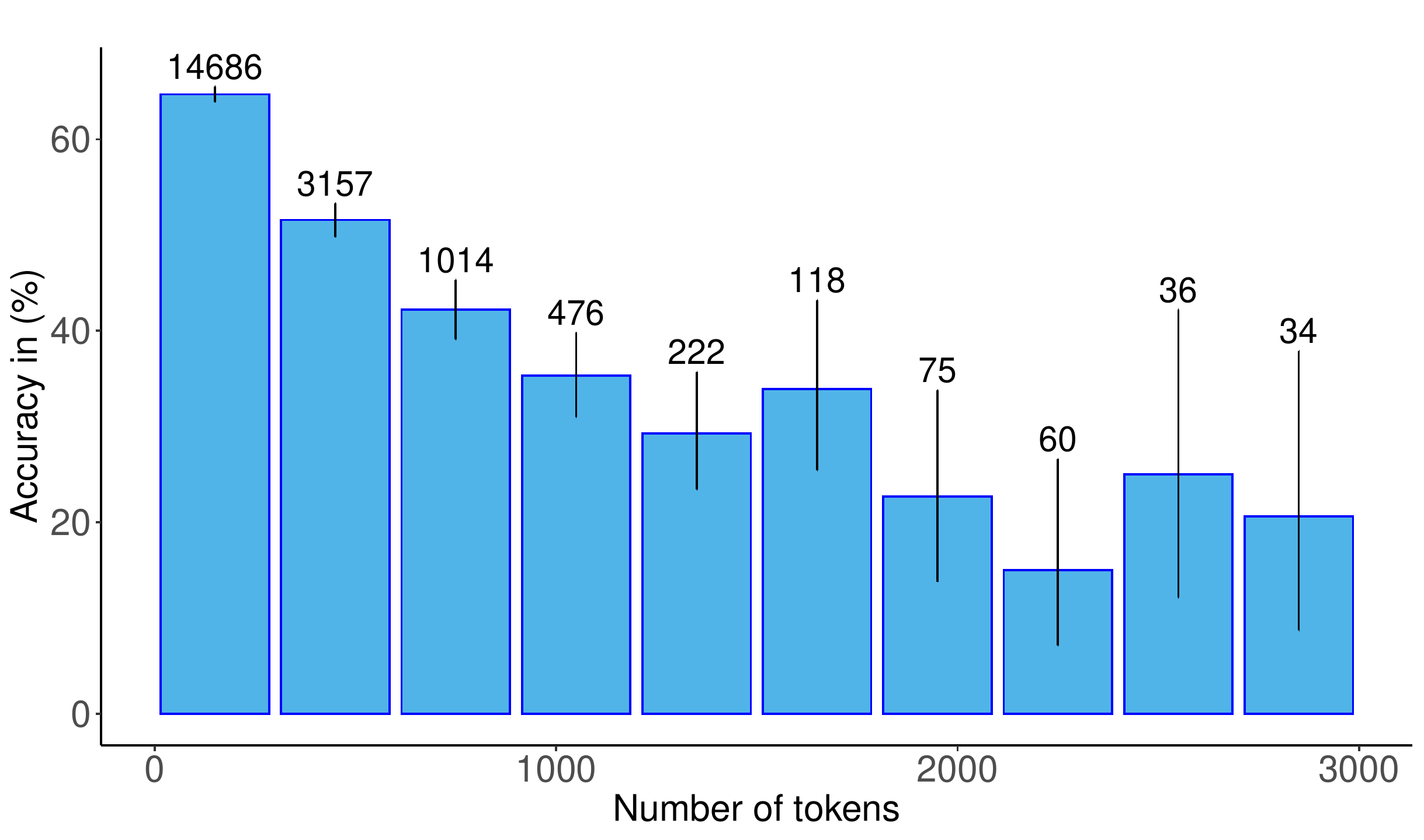}
  %\caption{Precision}
  %\label{fig:sub1}

\caption{Performance of code correct with increasing number of tokens in the file with 95\% confidence interval}
\label{tokenaccuracy}
\end{figure}

\smallskip             
\noindent{\underline{\emph{Performance vs. file length}}}
Most student programs are less than 1K tokens in length (though some are much longer). 
%Though there are some files with thousands of tokens, 95\% of the student program has less than 1000 tokens. 
It is reasonable to expect performance to decrease with (much) larger files; indeed, even the \taseg could struggle with large files, since even abstracted version of these can have hundreds of tokens.   
Figure~\ref{tokenaccuracy} shows how (Top-1 accuracy) performance decreases with length. For simplicity, we bucketed the samples by length, and show average performance and confidence intervals for each bucket.
We can see that our pipeline achieves a peak performance of around 63\% accuracy for files with less than 300 tokens, while accuracy decreases to ca. 20\% around 3000 tokens.
Note that the confidence interval increases with length, because there are fewer and fewer samples in our data (bucket size is indicated above each bar). It should be noted that files under 1000 tokens, our top-1 accuracy is around 58\%, which compares favourably with other approaches. 

To observe the performance of \taseg with increased file length, we prepared ten buckets with different ranges of token counts. Each bucket consists of 10K examples.
Table~\ref{blk} shows that the accuracy of \taseg significantly declines with file length. We also report the MRR if we do not apply \taseg. The MRR does not change much with increased file length. Therefore, we can infer that our model works consistently on all the errors except the one due to imbalanced braces.  

\begin{table}[]
\caption{Performance of \taseg vs. file length}
\label{blk}
\centering
\begin{tabular}{|C{2cm}|C{1.5cm}|C{1.5cm}|C{1.5cm}|C{1.5cm}|C{2cm}|}
\hline
\textbf{Token Range} & \textbf{MRR (overall)} & \textbf{Block Error} & \textbf{Solved by \taseg} & \textbf{Accuracy (in \%) of \taseg} & \textbf{MRR (without \taseg)} \\ \hline
1-100                & 0.66         & 2426                 & 2124                        & 87.55                                          & 0.45                            \\ \hline
101-200              & 0.62         & 2403                 & 1787                        & 74.36                                          & 0.44                            \\ \hline
201-300              & 0.56         & 2491                 & 1536                        & 61.66                                          & 0.41                            \\ \hline
301-400              & 0.52         & 2417                 & 1200                        & 49.64                                          & 0.40                            \\ \hline
401-500              & 0.47         & 2492                 & 1024                        & 41.09                                          & 0.37                            \\ \hline
501-600              & 0.44         & 2448                 & 737                         & 30.10                                          & 0.36                            \\ \hline
601-700              & 0.39         & 2343                 & 519                         & 22.15                                          & 0.34                            \\ \hline
701-800              &0.36              &2178              &359                             &                                           16.48     &0.32                                 \\ \hline
801-900              & 0.33             & 2362                     & 284                            &                                               12.02 &0.31                                 \\ \hline
901-1000              &0.31              & 2183                     & 149                            &6.08                                & 0.29                                \\ \hline

\multicolumn{6}{l}{Note: * we have 9,130 examples for 701-800 tokens and 9,052 for 901-1000 tokens.} 
\end{tabular}
\end{table}

%\toufique{keep any one from below}
  
%\begin{figure}[htb]
%\centering
%
%  \centering
%  \includegraphics[scale=.4]{tokenaccuracy_percent.pdf}
%  %\caption{Precision}
%  %\label{fig:sub1}
%
%\caption{Performance of code correct with increasing number of tokens in the file with confidence interval}
%\label{tokenaccuracy1}
%\end{figure}         

\begin{figure}
  \centering
  \begin{tabular}{@{}c@{}}
    \includegraphics[width=.7\linewidth,height=135pt]{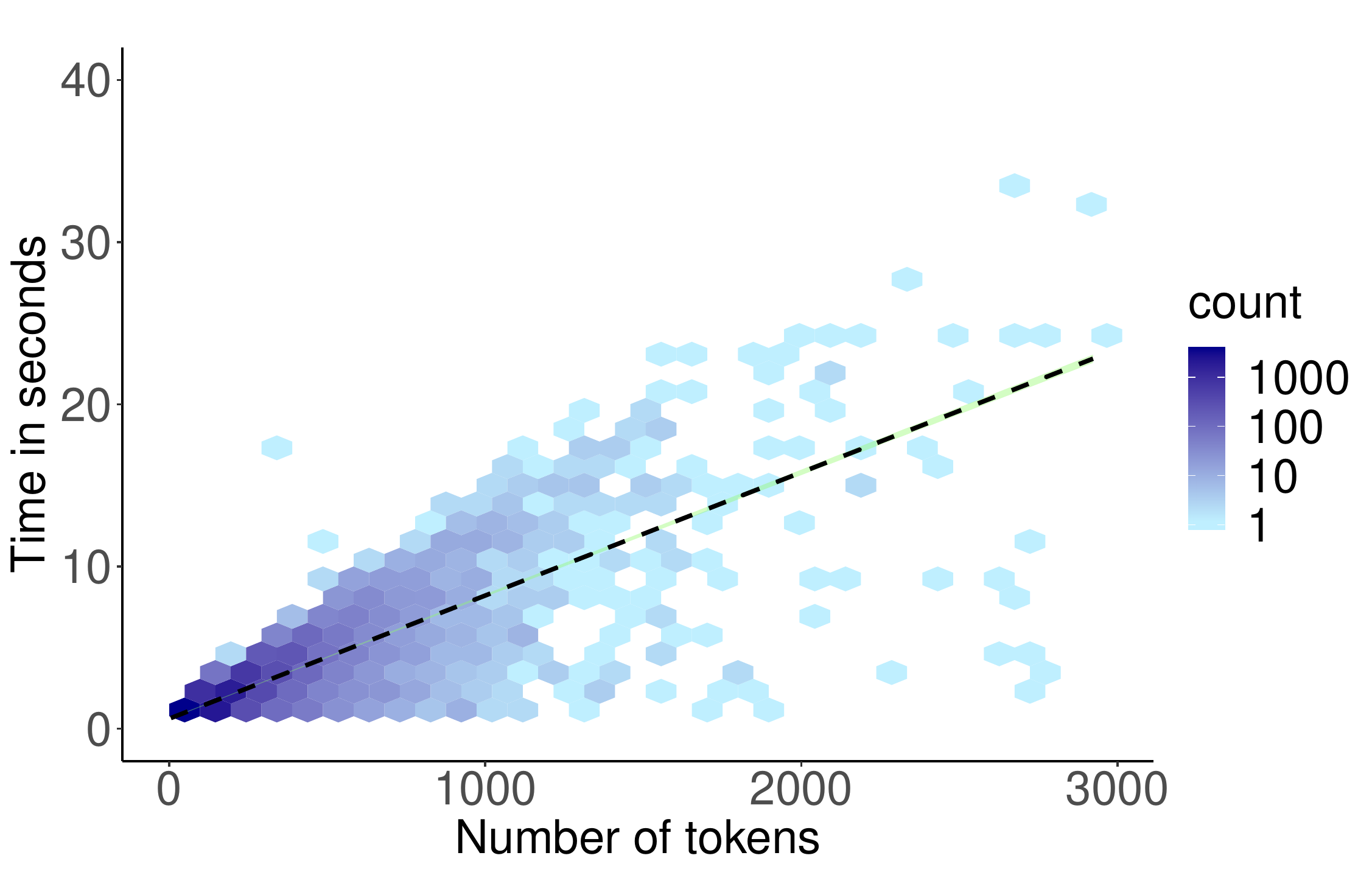} \\[\abovecaptionskip]
    \small (a) Fixable
  \end{tabular}

  \vspace{\floatsep}

  \begin{tabular}{@{}c@{}}
    \includegraphics[width=.7\linewidth,height=135pt]{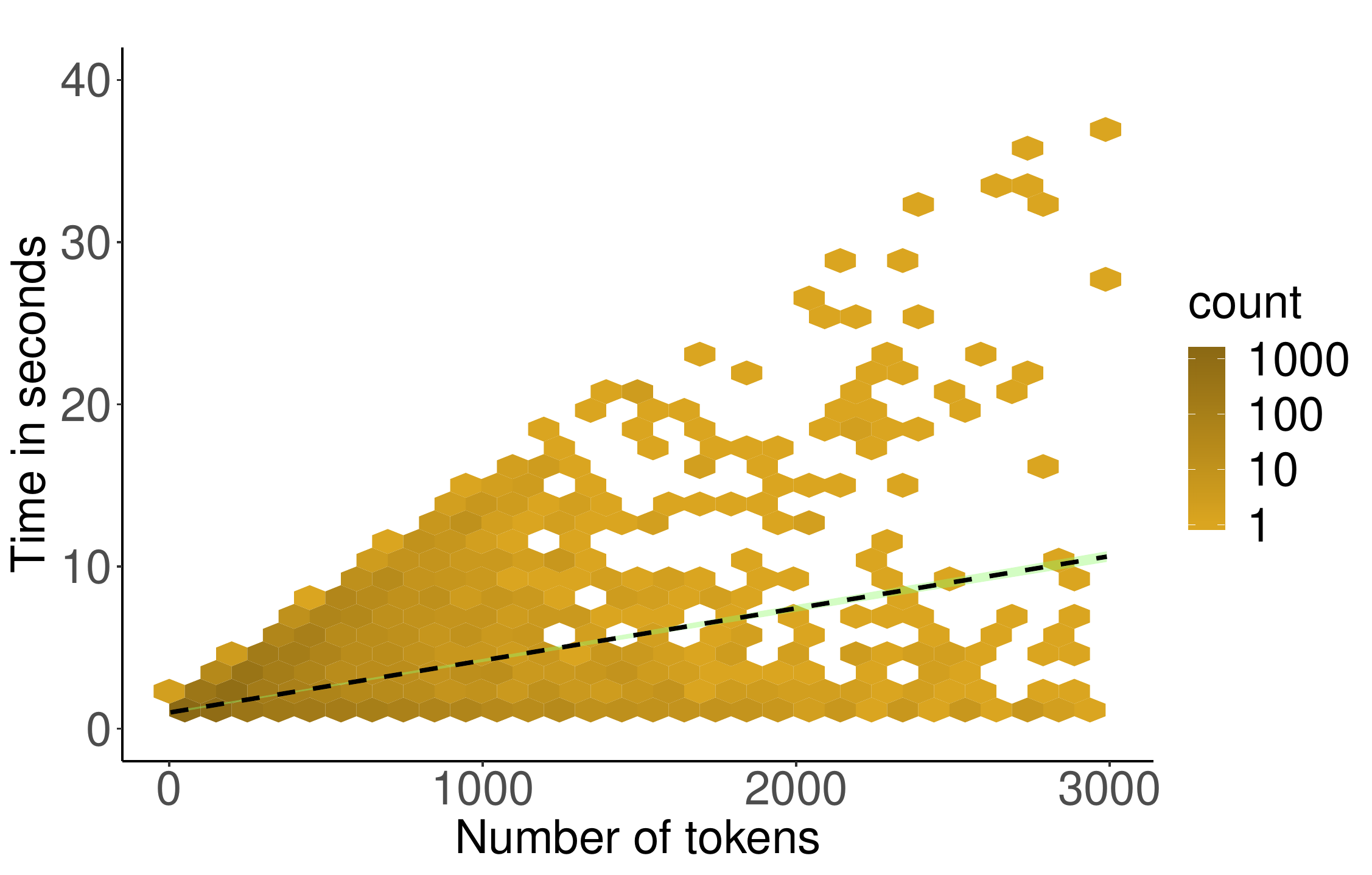} \\[\abovecaptionskip]
    \small (b) Non-Fixable
  \end{tabular}

  \caption{Processing time vs. number of tokens in the file}\label{timelength}
\end{figure}

\smallskip             
\noindent{\underline{\emph{Time vs. Length.}}}
Our biggest performance overhead is the DNN computation time, especially since we use two separate models; we can expect our pipeline to take longer for bigger files. 
%We use two DNN models to fix the student program errors. These DNN models can be slow sometimes. Besides, we apply a DNN model on each segment, which may slow down the whole fixing process.
To assess this, we measured performance on a random sample of 20,000 files, and show separate plots for cases
where we succeeded and failed in Figure~\ref{timelength}. Note that these are scatter plots that additionally signal the prevalence of datapoint buckets with their color gradient. 

Immediately evident is the flatter slope for the failing cases, with many failing quickly: 
these usually fail earlier in our pipeline---either \taseg fails to properly balance and nest the input source code,  or there
other errors that inhibit fragmenting of the code, so we abort before getting to \taparse. Figure~\ref{timelength} also  shows that  the processing time generally increases with the number of tokens.
Even so, most files are processed fairly quickly. Our median repair time is around 1.5 seconds, which is about 10\% of the median repair time reported in the Blackbox dataset (gathered from actual human-generated fixes, see Table 1~\citep{brown2017novice}), %\toufique{they reported 11 types of syntactical error with 13-1000s median time. The class with 13s repair time is not the biggest one. 10\% is not reflecting the time efficiency}
suggesting that we could provide timely help to  students quite often.  
%students with assistance before a Google search completes, or a friend responds to a plea for help.
In all, we process 95\% of files in under 10 seconds. 
%\prem{two plots, one for each of correct an dincorre} 

%\caption{Processing time vs. number of tokens in the file}
%\label{timelength}

%

\smallskip             
\noindent{\underline{\emph{Comparison with DeepFix~\citep{gupta2017deepfix}}}}
DeepFix, which uses an RNN seq2seq (with a single attention head) translation model, also works on student programs and repairs syntactic errors in C with 27\% top-1 accuracy~\citep{gupta2017deepfix}. The programs in  Deepfix's dataset range in size from 100-400 tokens, making them substantially much smaller than those in  Blackbox.

Typically, DeepFix considers an erroneous program of a few hundred tokens;
precisely predicting the (fixed) target sequence this long is very challenging for neural networks. 
Natural language translation (which was the original intended application of seq2seq neural models) 
rarely need to handle such long sequences, since they work a sentence at a time. 
%None of the current state-of-the-art machine translation models can support that long translation. 
To combat this problem, they encode the problem representation with line number. That means instead of generating all the program tokens, their model outputs the defective statement with the line number, which is an excellent idea of representing the error in the program. However, there is no straight forward way to
directly compare our work with DeepFix's current learned model, because of 3 reasons.

\begin{enumerate}

\item Deepfix is trained on C programs, which has a different syntax. Therefore, the reported accuracy is not directly comparable. 
Therefore, we trained the (publicly available) DeepFix implementation\footnote{See \url{https://bitbucket.org/iiscseal/deepfix/src/master/}} on our Java dataset. 

\item In DeepFix, the authors applied mutations to generate training examples because of limited programs from 93 different problems. They introduced five mutations on each program and applied five-fold cross-validation. In each fold, they have an average of 200K mutations on their training set. They iteratively resolve the errors from the program one at a time. Since we have a 200K real student program with a single error, we train DeepFix with real data instead of mutated ones.

\item As mentioned earlier, the length of the programs ranges from 100-400 tokens for DeepFix. Since DeepFix is trained on all the tokens of a program, there must be some upper limit on token counts. This limitation is fundamental to the structure of all state-of-the-art machine translation models. Moreover, the training typically gets harder for longer sequences. There are a significant amount of Java programs with more than 400 tokens. It would be challenging for a model trained with limited tokens and explicit line number to repair the error in longer sequences. Therefore, we trained two DeepFix models: DeepFix Short (DF Short) and DeepFix Long (DF Long) to compare with our one. DF Shorts is trained on examples  up to 400 tokens long, and DF Long is trained with examples up to 800 tokens long. We also prepared ten buckets of programs with token counts  upto 1000 tokens for validation. We ensured that none of the examples from the test set is present in the training set. Finally, we evaluate three models (\taseg + \taparse , DF Short, and DF Long) with the test set. Note that we applied these models on files with single syntactical error.             

\end{enumerate}   

\begin{figure}[h]
\centering

  \centering
  \includegraphics[scale=0.45]{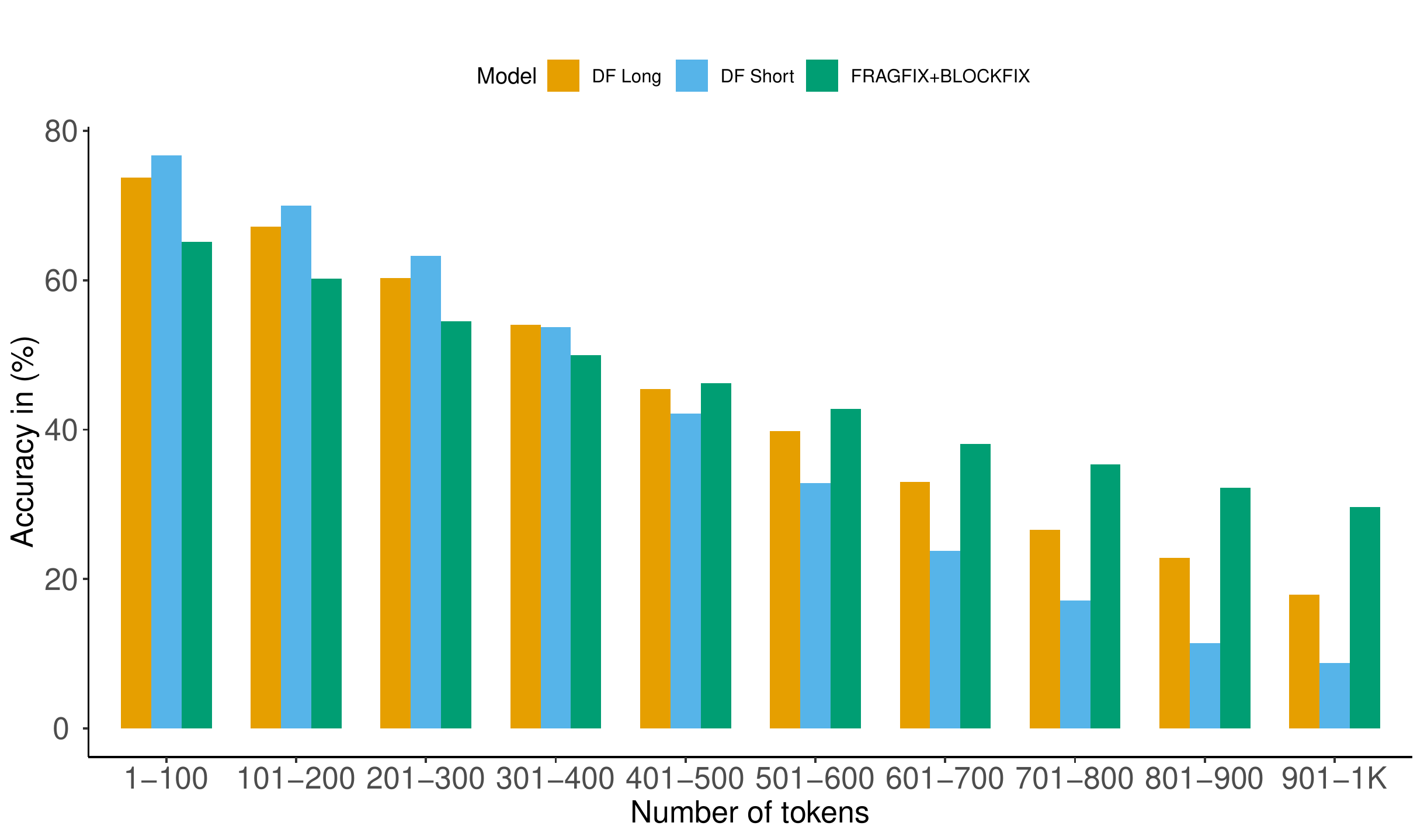}
  %\caption{Precision}
  %\label{fig:sub1}

\caption{Performance of three models on programs with different ranges of token-count}
\label{deepfix1}
\end{figure}

Figure~\ref{deepfix1} shows that DF Short and DF Long perform a bit better than our model (\taseg + \taparse) until 400 tokens. After 400 tokens, our model surpasses the other two models and then gradually becomes dominant. The accuracy of all the model decrease with increased file length. DF Short worked extremely well with the shorter programs (around 76\% top-1 accuracy for 1-100 tokens) but fails poorly with longer ones (8.8\% for 901-1,000 tokens). It is quite obvious that if the error occurs after 400 tokens, the corresponding line number may not be in the output vocabulary. Without the line number, it is not possible the locate the error correctly. DF Long performs almost similar to DF Short; however, it performs well after 400 tokens and achieves 17.4\% top-1 accuracy for 901-1,000 tokens. We can conclude that DeepFix models will fail to resolve minor or prevalent errors if the error located outside the token-range of training data. Besides, locating an error in the longer program is inherently difficult than to locate in shorter programs.      

Though our model is not the best one for shorter programs (around 65\% top-1 accuracy for 1-100 tokens), it achieves higher accuracy on the longer ones (30\% top-1 accuracy for 901-1,000 tokens). We applied our model on even longer sequences (10,000 examples of 1,000-3,000 tokens) and we still achieves 26\% top-1 accuracy. As discussed earlier, in our approach, we divide the program into smaller fragments and try to repair each error locally. Therefore, we could solve more trivial errors in the longer program compared to other models, and even for those errors, there is no need to retrain \taparse or \taseg. 
However, there are several reasons for getting lower accuracies in the shorter programs. We generate AST on the \taparse and introduced about 20 common mutations.
Moreover, because of segmentation, we lose some information (i.e., the position of the token in the real program). DeepFix model performs better at finding misspelled keywords, and it extracts more than 20 mutations from the training data. On the other hand, our model does not work with misspelled keywords. Since we do not have positional information, it is not very easy for our model to work on this. Consider the following statement with a syntactic error from the student program.

\begin{smcodetabbing}
\textcolor{red}{imkport} slp.core.util.Pair;
\end{smcodetabbing}  

Let us consider ``imkport" is not present in the input vocabulary of both \taparse and DeepFix model. Therefore, both models will encode it with ``$<$unk$>$". Now, \taparse does not have any positional information; it assumes two identifiers (all ``$<$unk$>$" are identifiers) are placed side by side. In that case, our model may try to separate two identifiers by placing a period between two identifiers. Moreover, \taparse fails to identify the root node of the AST: ``ImportDeclaration". 

\begin{smcodetabbing}
\textcolor{red}{imkport.} slp.core.util.Pair;
\end{smcodetabbing} 

On the other hand, DeepFix can fix this. In most of the cases, the first few lines of java programs start with keyword ``import". 
DeepFix quickly learns this pattern using the line number.         

\begin{smcodetabbing}
line-number1 \textcolor{green}{import} slp.core.util.Pair;
\end{smcodetabbing}

\begin{figure}[htb]
\centering

  \centering
  \includegraphics[scale=0.45]{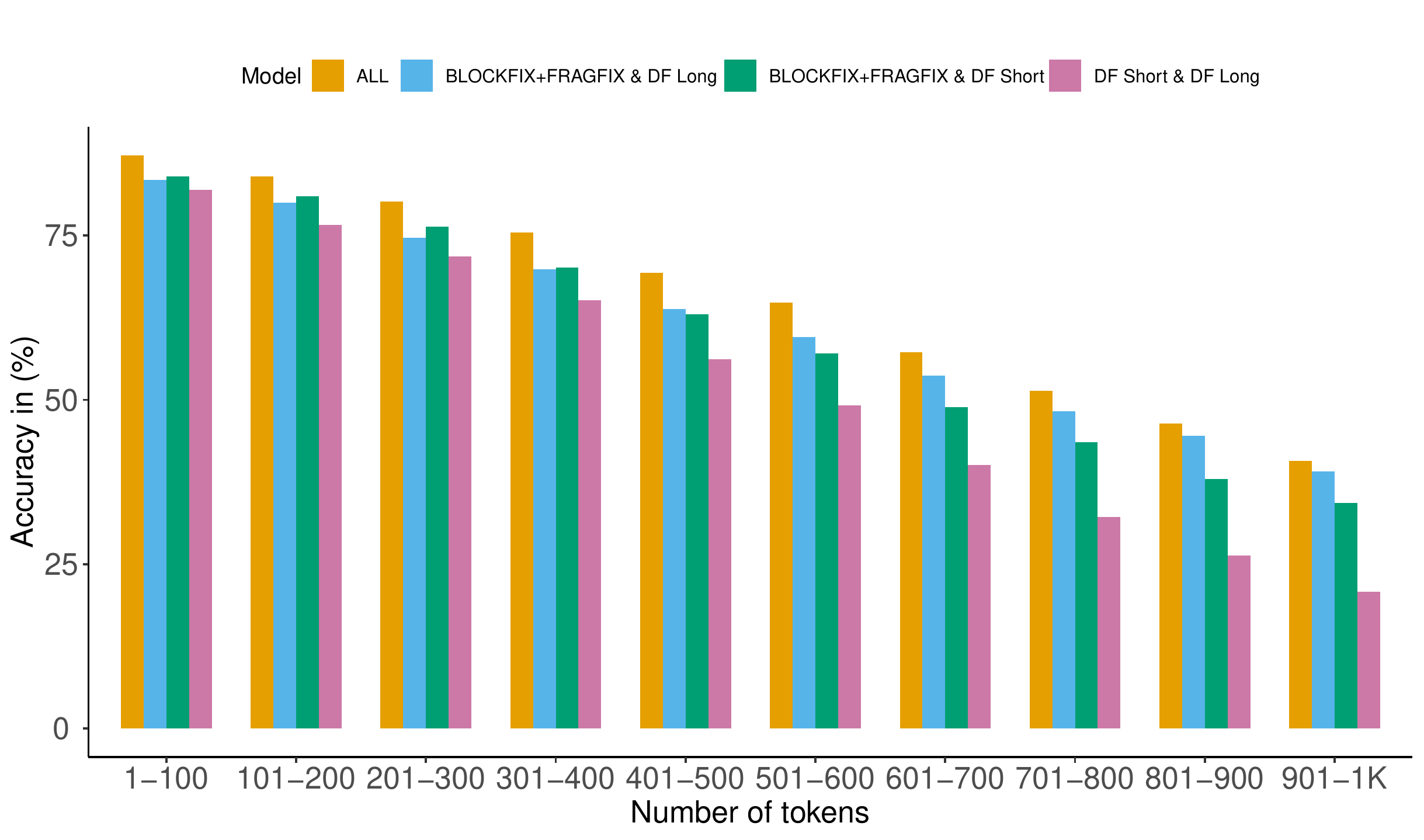}
  %\caption{Precision}
  %\label{fig:sub1}

\caption{Performance of blended models on programs with different ranges of token-count}
\label{deepfix2}
\end{figure}

DeepFix~\citep{gupta2017deepfix} is also reasonably language agnostic; on the other hand, while it can repair code, it's not a "lenient" parser that can parse fragmentary, and wrong code. We can produce a tree for downstream use, e.g., from a fragment of case statement or an statement with ellipses (``...'').

\smallskip             
\noindent{\underline{\emph{Blending \taseg + \taparse and DeepFix~\citep{gupta2017deepfix}}}}
Finally, we explore a blended strategy. 
Do DeepFix and  \taseg + \taparse are repairing the same files? Since we have a fixed validation set, we try to blend three models to see how much top-1 accuracies we could achieve. 
We apply a simple heuristics to report the blending accuracy. If an error is corrected by at least one of the candidate models, we would consider that as a success. On the contrary, if all the candidate models fail to correct one specific error, we marked that as failure by the blending model.

Figure~\ref{deepfix2} presents the accuracies if we blend the models. If we consider all the three models, the top-1 accuracies range from 87\%-40\% for different token ranges. 
On the full dataset, 
we can achieve 77\% overall top-1 accuracy, which outperforms the individual performance of all the models. 
For any given erroneous code, we get 3 guesses, one from each model; using the compiler we can pick
a good fix. If the compiler rejects the result from a certain model, we can try another one. We also note that mixing
 DF Short and DF Long (prior work) doesn't provide the best results. 
Therefore, we can infer that there are a significant amount of errors (10\% of Blackbox dataset) are solved by our model, which the
DeepFix approach cannot handle.

%!TEX root=main.tex

\section{Related Work}
\label{relatedwork}
\smallskip             
\noindent{\underline{\emph{Island Parsing}}}
The main objective of the island parsing problem is to find ``islands" of structured content (\emph{e.g.,} code snippets) from
``water'' of unstructured data (\emph{e.g.,}  English descriptions). Since 
useful code snippets are often found in mixed English-code corpora in manuals, web sites, \emph{etc}, 
%these code snippets are useful and \sofo fragments are noisy because of mixed-up natural language and code fragments, 
island parsing can help programmers by  carving out useful bits of code. 
%using code snippets by isolating them from natural language. 
Moonen and Van Deursen
 introduced a grammar-based approach to solving island parsing problem~\citep{van1999building, moonen2001generating}. 
Synytskyy~\citep{synytskyy2003robust} demonstrate
the use of  this approach for dealing with ASP fragments, which mix comments, HTML, and Visual Basic. 
Bacchelli \etal applied two approaches to solve island parsing problem: generalized LR (SGLR) and Parsing Expression Grammars (PEGs)~\citep{bacchelli2017mining}. Rigby \etal did not separate the code snippet from \sofo fragments; instead, they applied a set of regular expressions to approximate the java construct, e.g.,  qualified terms, package names, variable declarations, qualified variables, method chains, and class definitions~\citep{rigby2013discovering}.  While this is a powerful approach, current methods depend on hand-crafted grammars. Our approach is rather more general, requiring just the availability of a parser that can produce ASTs. 
Though Island parsers might (See~\citep{bacchelli2017mining}, \S 6.3) be applicable to code with syntax errors, we are not aware of any prior work or benchmark where they were used to correct student code to compare our work against. 

A related line of work is \emph{partial program analysis}, which attempts to derive types and data-flow facts from 
incomplete programs~\citep{rountev1999data, dagenais2008enabling}. Most of these works in the
area of partial program analysis consider ``fragments" to be either complete files, or complete procedures, rather than the kinds of noisy bits we consider.  The one available tool, PPA\footnote{\url{http://www.sable.mcgill.ca/ppa/}} only works for Java 1.4 or 1.5. Our test set (and training corpora) include features from later releases (such as  Collections). 
We also note a considerable body of prior work in finding, using, and mining code examples from the web~\citep{holmes2005strathcona, thummalapenta2007parseweb, nasehi2012makes,ponzanelli2014mining}. Our work is generally complementary to this line of work.

\smallskip             
\noindent{\underline{\emph{Predicting the Type of Identifiers}}}
Dagenais proposed some predefined strategies to infer the types of identifiers, e.g., by using the type of the identifier on the other side of assignment operator~\citep{dagenais2008enabling}.
Alexandru \etal propose that DNN is not very good at tokenizing source code, but it is highly capable of recognizing token types and their relative locations in a parse tree~\citep{alexandru2017replicating}.
Raychev \etal applied statistical inference model for inferring types for JavaScript~\citep{raychev2015predicting}. Hellendoorn \etal~\citep{hellendoorn2018deep} use DNN to predict the types. While these works use the implementation to infer types, Malik \etal extract type information from natural language descriptions (comments, identifiers) \citep{malik2019nl2type}. However, none of these machine learning-based approaches were applied specifically to inferring the type of incomplete code fragments; they were trained and tested on complete source code files. 

\smallskip             
\noindent{\underline{\emph{Fixing Compile Errors}}}
Mesbah\emph{ et al.} describe DeepDelta, which fixes mostly identifier name related errors, not syntax errors. 
DeepDelta was developed and tested on code that led to build errors, all from professional developers at Google. 
The authors also assume that precise knowledge of the location to be fixed is available ~\citep{mesbah2019deepdelta}, which we do not.

Gupta \etal applied reinforcement learning to a very similar dataset \citep{gupta2018deep}, 
reporting 26.6\% accuracy of their tool (RLAssist). Bhatia \emph{et al.} achieved slightly higher accuracy than Deepfix and RLAssist on repairing student code (31.69\% accuracy)~\citep{bhatia2016automated}. However, these numbers are difficult to compare across datasets; e.g., Bhatia \etal's dataset consists of solutions to just 5 different programs, which is considerably less diverse than BlackBox (which collects data from all users of BlueJ, not just ones doing particular homeworks). The programs in~\citep{bhatia2016automated} are also relatively small, ranging from ca. 40 to 100 tokens.
Santos \textit{et al.} used the BlackBox dataset, and were able to fix almost half of instances of student code with single syntax errors \citep{santos2018syntax}; as reported earlier, we exceed their MRR performance, and can also fix
programs with more than one error.

\noindent{\underline{\emph{Deep-learning for Code Repair}}} There is considerable interest in applying deep
learning to the problem of code repair. Typically, such work uses a large dataset of bug-fixing commits to train seq2seq type models~\citep{tufano2019learning,chen2019sequencer,DLFix2020,ding2020patching,lutellier2020coconut}, or to find relevant repairs for patching~\citep{white2019sorting}. Translation models that use tree to tree (rather than seq2seq) have also been proposed~\citep{chakraborty2018tree2tree,DBLP:journals/corr/abs-1810-00314}. These approaches
are mostly not aimed at syntax errors, but rather at semantic errors exposed by failing tests. However, Deepfix~\citep{gupta2018deep} is relevant to our approach, and we have done an extensive comparison above. 

All the above mentioned Automatic Program Repair (APR) tasks have a fault localization step. APR is inherently a harder problem because it is often unclear where the semantic error is. The success of these tools depends a lot on the fault-localizer. Different APR techniques use different fault-localizers. Most of the papers use the perfect fault-localizer to have a fair comparison. We can infer that the reported performance will not be the same if the authors would use some real-world, imperfect fault-localizer. For our task, we do not need any fault-localizer. \taseg \& \taparse are together capable of localizing and fixing the error. %Furthermore  most APR approaches such as GenProg~\citep{le2011genprog} and Prophet~\citep{long2016automatic} assume that the old version is syntactically correct
%and can be compiled and executed, or subject to static analysis. Our work is aimed directly at programs that have syntax errors.

We note that many of these ``semantic" models are not applicable to our setting. For one, most APR approaches, including GenProg~\citep{le2011genprog}, Prophet~\citep{long2016automatic}, as well as tree to tree neural models (Chakraborty et al., 2018a,b), assume that the “old” (buggy) version of the code is at least syntactically valid and can be parsed/compiled/executed, and even subjected to static analysis. Needless to say, this is inapplicable to our setting. Furthermore, most deep learning-based work are specifically trained to repair semantic errors~\citep{tufano2019learning,chen2019sequencer,DLFix2020,ding2020patching,lutellier2020coconut}, or to find relevant repairs for patching~\citep{white2019sorting}. Without training on syntactic errors, such models are highly unlikely to be useful on the latter. DeepFix~\citep{gupta2018deep}, however, does target syntactic errors and is thus precisely relevant to our approach, and correspondingly we compared extensively with that tool. 
%Naturally, one could imagine retraining some of the former models (at least, the ones not requiring test cases or parsing) on a dataset of syntactic errors, but there is little reason to assume that they would work any better in our setting (viz., fixing syntax errors, rather than bug-patching) than DeepFix; we do indeed have a comparison with DeepFix.

%\prem{Todo: Check Bhatia et al. and RL ASSist to see what's what}.  

%\smallskip             
%\noindent{\underline{\emph{Attention is all you need}}}
%\prem{Transformers are considered best so far}
%We applied the Transformer based DNN model for all of our experiments~\cite{vaswani2017attention}. Transformer based model uses multi-head self-attention to collect information from different parts of the sequence and remove the recurrence part. It outperforms all the state-of-the-art translation model with or without attention mechanism~\cite{bahdanau2014neural,hochreiter1997long,chung2014empirical}. 

%\smallskip             
%\noindent{\underline{\emph{Fixing student Program}}} 

%island parsing
%deepfix, 
%synfix, VERY CLOSE IN PERFORMANCE !!! but trained on data of the same assignment. 
%sense \& syntax
%when we say best-class we mean: when not previously training on homeworks of exactly teh same type, %for syntax errors. this is also matters. 

%attention is all you need/BERT 

%!TEX root=main.tex

\section{Discussion \& Future Work}
\label{sec:discussion}
\subsection{Threats \& Caveats}  
Despite the observed performance, some caveats apply. 
For the \sofo parsing task, even with nearly 90\% accuracy for the combined approach, developers will still have to deal with erroneous repairs. Although we did not run experiments with developers, we can expect that, if used within an IDE, features like syntax-directed indenting should make it fairly easy for developers to assert whether the pasted-in AST is indeed correct. Our manual assessment relied on a random sample; the confidence interval reported (\S~\ref{sec:sofoparse}) gives a sense of how the actual performance might vary.

For the student code syntax correction task, our top-1 accuracy estimate (matching the exact fix produced
by the student) is based on a very large random sample, and is thus likely to be close to the true value. Although higher than previous work, we still reach only 56\% top-1 accuracy; thus, suggested repairs may still be incorrect, either syntactically or semantically. To ensure that the fix is good syntactically, it would be prudent to apply the fix and run the compiler or a parser, as a check (which can be done automatically) before offering the suggested fix to the user. Semantic correctness of the suggest repair (or at least equivalence to what student intended) is much more problematic to determine, and can only be assessed with test cases or invariants provided by the instructor. 

Our performance on the lenient typing task is good for popular types, but clearly declines with decreased
training data availability. While we hope to improve the performance in future work (see below), the type annotations
especially for less common types would need review by the developer if used in an IDE.

\subsection{Future Work}
There are several interesting directions for extensions of this work.
Our lenient parser uses indirect supervision on noised data and was not trained on student or \sofo data.
However, there is a lot of student data available, which should provide a more precise signal, if relatively less training data. In that light, it is entirely reasonable to see our current setting as a form of pretraining and additionally fine-tune our model on e.g. real student data, using the true fixes as targets.
This might improve performance in ways attenuated to student data specifically.
  
Since our lenient parser provides an actual repaired parse tree, and not just a suggested edit, 
there is also an opportunity for each suggested fix to provide some pedagogical value and/or explanation
as to \emph{why} some token(s) should be added, removed, or changed. This is a promising direction
we hope to pursue. 

Lenient typing performance is currently constrained, first because of limited data, and secondly because
the vocabulary is limited for the input embedding layer. As noted by Malik \etal~\citep{malik2019nl2type} quite a bit of type information is carried in the identifier names; so we
believe that approaches like~\citep{babii2019modeling, karampatsis2020big}, which intelligently
decompose identifiers into constituent sub-tokens based on co-occurrence frequencies, can enhance
performance for the typing task, since the compositions of identifiers can be used predict their types.  
This will require recasting the typing task as translation (rather than tagging)
since input and output lengths won't match anymore. Finally, we also believe that both lenient parsers
and types for domain-specific IDEs (such as Android Studio for Android, or Visual Studio for .NET) could benefit from training on large volumes of code rich in specific APIs of interest to the target audience.

%!TEX root=main.tex

\section{Conclusion}
\label{sec:conclusion}
We have described an approach to processing  (parsing \& typing)
incomplete and erroneous code,
from students and \sofo. We generate large volumes
of  training data for parsing \& typing erroneous code by starting with
code which syntactically correct, and well-typed, which can be parsed and typed with a standard parser, and then fragmenting and injecting noise into this data to train a lenient parser and typer. 
We use a parsing-as-translation approach, based on the state-of-the-art
Transformer model, while using a tagging approach for typing. 
To deal with the long-distance dependencies
of source code, we first segment the code into fragments, using statement
delimiters and nesting via curly braces. Since code could have improper nesting,
we train a separate model to fix missing or extra nesting structures. 
This pipeline, consisting of  \taseg and \taparse, performs
better on the large and diverse BlackBox dataset than previous work. It also performs 
well for StackOverflow fragment parsing, and has some degree of success
on the typing task. In future work, we hope to pursue further improvements on the typing task,
and seek integration with an IDE, to help fix errors, and also to help paste-in
code from StackOverflow. 

Finally, we will make our implementation, and some of the data available at \url{https://doi.org/10.5281/zenodo.3374019}. 
The BlackBox data is \emph{not} redistributable, and must be explicitly requested from the authors~\citep{brown2014blackbox}.

\bibliographystyle{spbasic}      
\bibliography{acmart.bib, sample-base.bib}

\end{document}